\documentclass[nopacs,notitlepage,preprintnumbers,amsmath,amssymb,nofootinbib,superscriptaddress]{revtex4-1}
\usepackage[dvips]{graphicx}
\usepackage{amssymb}
\usepackage{amsmath}
\usepackage{graphicx,color}

\usepackage{enumerate}
\usepackage{mathtools}
\usepackage{stmaryrd}

\usepackage{epstopdf}

\usepackage{textcomp}
\usepackage{gensymb}

\newcommand{\moy}[1]{\left\langle #1 \right\rangle}

\newcommand{\TT}[0]{\boldsymbol{T}}

\newcommand{\ex}[1]{\mathrm{e}^{#1}}

\newcommand{\nn}[0]{\hat{\boldsymbol{n}}}

\newcommand{\rr}[0]{\boldsymbol{r}}
\newcommand{\qq}[0]{\boldsymbol{q}}

\newcommand{\ii}[0]{\mathrm{i}}
\newcommand{\dd}[0]{\mathrm{d}}
\newcommand{\Pe}[0]{\text{Pe}}

\newcommand{\rotop}[0]{\boldsymbol{\mathcal{R}}}

\newcommand{\ee}[0]{\boldsymbol{e}}

\newcommand{\QQ}[0]{\mathbf{Q}}
\newcommand{\tf}[1]{\widetilde{#1}}

\definecolor{darkblue}{rgb}{0,0,0.6}
\definecolor{darkred}{rgb}{0.6,0,0}
\definecolor{forestgreen}{rgb}{0.13,0.55,0.13}
\usepackage[colorlinks=true,urlcolor=darkblue,citecolor=darkblue,linkcolor=darkred]{hyperref}

\begin{document}

\title{Closures of moment expansion of anisotropic active Brownian particles}
\author{Timoth\'ee Gautry}
\affiliation{Sorbonne Université, CNRS, Physical Chemistry of Electrolytes and Interfacial Nanosystems (PHENIX), Paris, France}
\affiliation{Sorbonne Université, CNRS, Laboratoire Jean Perrin, LJP, Paris, France}
\affiliation{Sorbonne Université, CNRS, Inserm, Institut de Biologie Paris-Seine, IBPS, Paris, France}

\author{Maxime Deforet}
\affiliation{Sorbonne Université, CNRS, Laboratoire Jean Perrin, LJP, Paris, France}
\affiliation{Sorbonne Université, CNRS, Inserm, Institut de Biologie Paris-Seine, IBPS, Paris, France}

\author{Pierre Illien}
\affiliation{Sorbonne Université, CNRS, Physical Chemistry of Electrolytes and Interfacial Nanosystems (PHENIX), Paris, France}

\date{\today}

\begin{abstract}
We study analytically the dynamics of anisotropic active Brownian particles (ABPs), and more precisely their intermediate scattering function (ISF). To this end, we develop a systematic closure scheme for the moment expansion of their Fokker–Planck equation. Starting from the coupled evolution of translational and orientational degrees of freedom, we derive equations for the density, polarization, and nematic tensor fields, which naturally generate an infinite hierarchy of higher-order moments. To obtain explicit solutions, we investigate truncation strategies and analyze closures at different orders. While the closure at lowest order yields Gaussian dynamics with an effective translational diffusion, closures at higher orders incorporate orientational correlations and reproduce non-Gaussian features in the ISF. By confronting these approximations with exact solutions based on spheroidal wave functions and with Brownian dynamics simulations, we identify their range of validity in terms of P\'eclet number, wavenumber, and observation timescales.  An advantage of this method is its ability to yield approximate yet explicit expressions not only for the ISF but also for polarization and nematic fields, which are often neglected but relevant in scattering experiments and theoretical modeling. Beyond providing a practical guide to select the appropriate closure according to the spatiotemporal regime, our framework highlights the efficiency of moment-based approaches compared to exact yet implicit formulations. This strategy can be systematically extended to more complex situations, such as propulsion switching, confinement, or external fields, where functional bases for exact solutions are generally unavailable.
\end{abstract}

\maketitle

\tableofcontents

\section{Introduction}

Active particles are encountered at many scales, from nanoscopic to macroscopic, both in biological and artificial systems. Their hallmark is their capacity to take up energy from their environment and to convert it into mechanical work in order to `self-propel'~\cite{Bechinger2016}. In  experiments and in theoretical modeling, their state is typically described by two vectors: their position $\rr$, and the orientation of their self-propulsion $\nn$. These two degrees of freedom are generally coupled: orientation, via self-propulsion, obviously affects the evolution of the position, and, in turn, the orientational dynamics may depend on the location of the particle. From the point of view of statistical mechanics, this coupling is what makes the mathematical study of active particles challenging and worthwhile.

Among the different ways to model active particles, `active Brownian particles' (ABPs) have become central during the past decades~\cite{Romanczuk2012,Solon2015a}. In the simplest version of this model, the orientation $\nn$ simply performs overdamped Brownian motion on the unit circle or the unit sphere, for ABPs in 2D and 3D respectively. The orientation of the particle typically has exponential correlations, i.e. $ \langle \nn(t)\cdot \nn(0) \rangle \sim \ex{-(d-1)D_r t} $, where $D_r$ is the rotational diffusion coefficient, and $d=2, 3$ is the spatial dimension. The position $\rr$ of the particle also obeys overdamped Brownian motion, to which a self-propulsion term $v\nn$ is added (where $v$ is homogeneous to a velocity and quantifies the intensity of self-propulsion).

In the ABP model, the mean-square displacement of the particle can be computed exactly. It typically shows a transition from a ballistic, persistent motion that results from self-propulsion, to an ultimate diffusive regime. The long-time effective diffusion coefficient of the particle reads $D_{0,\text{eff}} = D+\frac{v}{d(d-1)D_r}$. It has both a passive contribution, coming from translational Brownian motion with diffusion coefficient $D$, and an active contribution, coming from self-propulsion. This effective diffusion coefficient was for instance used to interprete pioneering experiments on active colloids~\cite{Howse2007}. Later on, the ABP model has been used to perform numerical simulations of interacting active particles, and to predict phenomena such as motility-induced phase separation~\cite{Fily2012,Redner2013,Cates2015}. 

To describe the dynamics of a system, computing the sole mean-square displacement (MSD) of active Brownian particles is not very specific, as alternative models, such as run-and-tumble particles (RTPs)~\cite{Martens2012}  or active Ornstein-Uhlenbeck particles (AOUPs)~\cite{Szamel2014} display the same mean-square displacements. Moreover, the MSD does not account for the fact that the statistics of the position of ABPs is non-Gaussian. Non-Gaussianity (which is ignored in simpler models, such as AOUPs) only appears when computing the full distribution, or at least moments of order three or higher. Up until a decade ago, very few theoretical studies addressed the problem of computing analytically the probability distribution function of the position of an active Brownian particle at arbitrary time. This seemingly simple problem appears to be quite challenging from the mathematical perspective. Indeed, self-propulsion couples orientational and translational degrees of freedom in a non-trivial way, in such a way that the Fokker-Planck equation obeyed by $P(\rr,\nn;t)$ (the probability to observe the ABP with orientation $\nn$, at position $\rr$ and  at time $t$) cannot be solved straightforwardly. However, significant progress was made in this direction. In 2016, Kurzthaler et al.~\cite{Kurzthaler2016} obtained an expression for the intermediate scattering function (ISF) of the ABP (i.e. the spatial Fourier transform of the marginal distribution associated with the joint position-orienation distribution, after integration over orientations). This expression is formally exact, but requires the numerical resolution of an eigenvalue problem, and is not fully analytically explicit. This result was extended to study ABPs with orientational resetting~\cite{Baouche2024}, to compute the first-passage time properties of ABPs~\cite{Baouche2025,Baouche2025a}, or to compute exactly the distribution of position of an ABP in a harmonic well~\cite{Caraglio2022}. Alternative studies resulted in asymptotic evaluations of the distribution of position in real space, both in the limit of short~\cite{Basu2018} and long times~\cite{Basu2019}. 

Finding explicit expressions to describe the dynamics of ABPs has then motivated the work of different groups during the past years, and our article contributes to this line of research.  With the exception of Ref.~\cite{Kurzthaler2016}, we emphasize that the case of anisotropic particles (i.e. whose mobility is described by a tensor rather than a scalar) has received little attention in analytical approaches. This adds another level of complexity to the model, since anisotropy generally makes the dynamics of the particle non-Gaussian, even in the absence of activity~\cite{Han2006}.

In this article, we propose to get explicit expressions of the ISF of anisotropic ABPs by performing a systematic moment expansion with respect to the orientational degree of freedom. We study different truncation orders and discuss the validity of this analytical scheme in the light of previous results. Although approximate, we believe that this approach provides an alternative way to study the dynamics of anisotropic ABPs, that may be of interest in situations where a fully explicit expression of the ISF is needed.

\section{Theoretical context}

\subsection{Fundamental equations and definitions}

\begin{figure}
\begin{center}
\includegraphics[width= 0.35\textwidth]{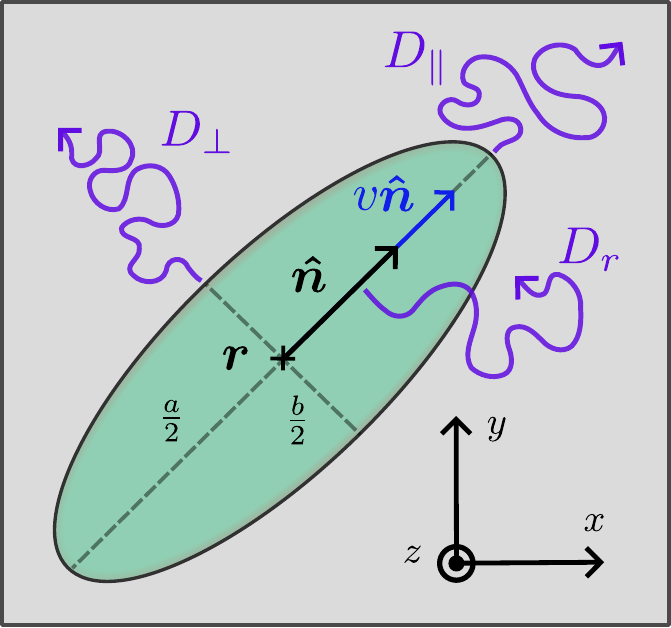}
\includegraphics[width= 0.4\textwidth]{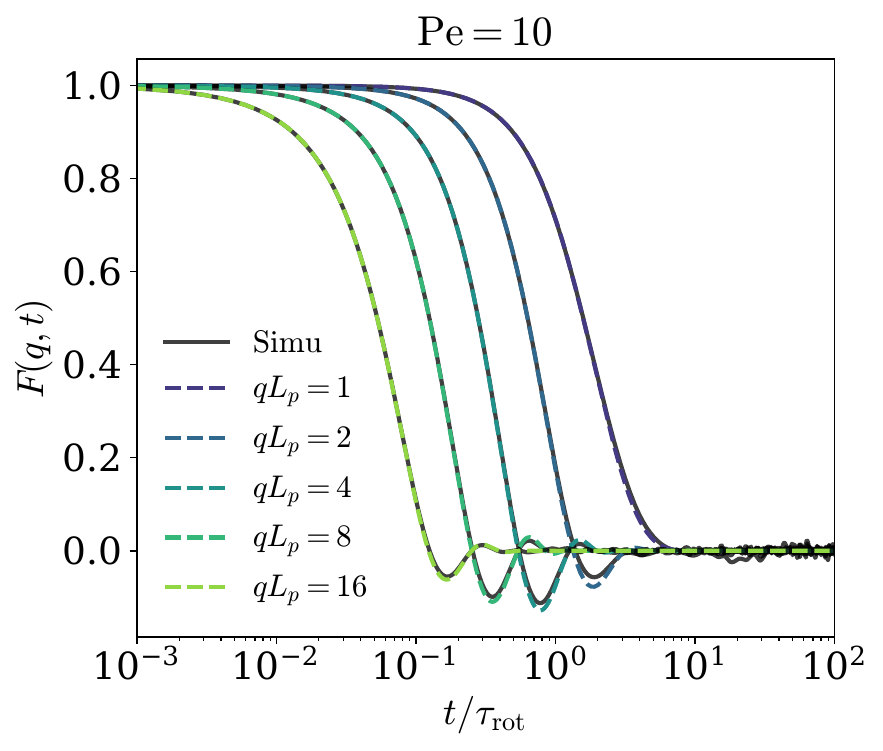}
\caption{Left: Sketch of the anisotropic active Brownian particle under study. Its position and orientation are denoted by $\rr$ and $\nn$ respectively. The longitudinal and transverse translational diffusion coefficients are denoted by $D_\parallel$ and $D_\perp$, while $D_r$ denotes the orientational diffusion coefficient. In the particular case of an ellipsoidal particle, $a$ and $b$ are its major and minor axes, respectively.  Right: intermediate scattering function, as a function of time, for different wavevectors, obtained from numerical simulations (see Appendix \ref{app:numerical} for details) and from the exact solution recalled in Appendix \ref{app:Kurzthaler_result}. Parameters: persistence length $L_p = 14$, aspect ratio $a/b = 3$, $v = 12$, $D_\parallel = 1.42$, $D_\perp = 1.16$, $D_r = 0.85$ ($\tau_{\text{rot}} = 1.17$).}
\label{fig:schema_ell}
\end{center}
\end{figure}

We consider an anisotropic active Brownian particle (ABP), in an unbounded space of dimension $d$ (with $d=2,3$) whose state at a given time $t$ is characterized by its position $\rr$ and its orientation $\nn$ (defined to be along the head-tail axis). It is self-propelled with a velocity $v$, along its orientation $\nn$, which fluctuates with a rotational diffusion coefficient $D_r$. The associated rotation time is defined as $\tau_\text{rot}=1/D_r$. We assume that the ABP is generally anisotropic. Therefore, its diffusivity is characterized by a tensor $\mathbf{D}$. As the particle self-propels along its head-tail axis, it is invariant by rotation around this direction, and therefore the diffusivity tensor $\mathbf{D}$ can be conveniently decomposed into a longitudinal and a transverse part. Its entries read:
\begin{equation}
D_{ij}\equiv D_{\parallel} n_i n_j + D_{\perp} (\delta_{ij} - n_i n_j).
\end{equation}
The coefficients $D_{\parallel}$ and $D_{\perp}$ are the longitudinal and transverse diffusion coefficients, respectively (Fig. \ref{fig:schema_ell}). The results we derive are valid for arbitrary $D_\parallel$ and $D_\perp$. For concreteness, in our numerical simulations, we will study the case where the particle is a prolate ellisoid, of major and minor axes $a$ and $b$, respectively. The relations between $D_\parallel$, $D_\perp$ and $D_r$ on the one hand, and $a$, $b$ on the other hand, are given in Appendix \ref{app:numerical}.

The probability $P(\rr,\nn;t)$ of finding the ABP at time $t$ at a given position $\rr$ and with a given orientation $\nn$ obeys the Fokker-Planck equation:
\begin{equation}
\label{eq:FPani}
\partial_t P(\rr,\nn;t)=\nabla\cdot(\mathbf{D}\cdot \nabla P) + D_r \rotop^2 P-v\nabla\cdot(\nn P)
\end{equation}
where $\nabla = \partial_{\rr}$,  and where $\rotop = \nn\times\partial_{\nn}$ is the rotational gradient operator\footnote{We recall here a few properties of the rotational gradient operator $\rotop = \nn \times \partial_{\nn}$~\cite{Doi1988, Dhont1996}. Its $i$-th component is written as $\mathcal{R}_i = \epsilon_{ijk} n_j \partial_{n_k}$, where $\epsilon_{ijk}$ is the Levi-Civita tensor. Although it is defined in 3D, it is can also be used in 2D: in this situation, $\rotop=\mathcal{R}_z \ee_z$, where $z$ is the direction perpendicular to the plane where $\nn$ belongs.  In this article, we use the following properties of the rotational gradient operator, which are valid for $d=2$ and $d=3$:
(i) Integration by parts can be extended to this operator, by writing $\int \dd \nn \;  f(\nn) \rotop g(\nn) = -\int \dd \nn \;  g(\nn) \rotop f(\nn) $, where the integrals are performed over the unit circle or the unit sphere; The boundary term vanishes because of periodicity;
(ii) When applied to a component of $\nn$, one gets $\mathcal{R}_i n_j = -\epsilon_{ijk}n_k$;
(iii) Given that $\epsilon_{ijk}$ is totally antisymmetric, one gets $\rotop \cdot \nn = 0$;
(iv) Finally, when applied twice, one gets $\rotop^2 \nn = -(d-1)\nn$ and $\rotop^2 \nn \nn = -2(d\nn \nn + \mathbf{1}) $.}. A rigorous notation for the propagator would be $\mathcal{P}(\rr,\nn|\rr_0,\nn_0;t)$, i.e. with a given initial condition. We will only consider the average of this propagator over the initial orientations and take the initial condition $\rr_0=0$ (in practice, this amounts to adding a term $\delta(\rr) \delta(t)$ in the right-hand side of Eq. \eqref{eq:FPani}).

\subsection{Quantities of interest}

The state of the system is described by the joint probability of observing the ABP at a given position and with a given orientation, denoted by $P(\rr,\nn;t)$, which obeys Eq. \eqref{eq:FPani}. The difficulty in the analysis of this equation essentially arises from the last term in the right-hand side: self-propulsion couples the evolution of the orientational and translational degrees of freedom. A quantity of interest is often the marginal distribution over positions only, which is defined as:
\begin{equation}
\label{eq:def_rho}
\rho(\rr,t) \equiv \int \dd{\nn}\; P(\rr,\nn;t), 
\end{equation}
where the integral is performed on the unit circle or the unit sphere, in dimensions 2 and 3 respectively. Another quantity of interest is the intermediate scattering function (ISF) of the ABP, denoted by $F(\qq,t)$, which is the Fourier transform of the positional probability distribution $\rho(\rr,t)$:
\begin{equation}
    F(\qq,t)  =\int \dd \rr\, \ex{-\ii \qq\cdot\rr} \rho(\rr,t).
\end{equation}
The reason for studying the distribution $\rho(\rr,t)$ or the ISF $F(\qq,t)$ is two-fold: (i) first, they both encode all the moments of the position of the ABP, and therefore account for the non-Gaussian behavior of ABPs; (ii) the ISF is the typical output of experiments based on differential dynamic microscopy (DDM)~\cite{Cerbino2008}, a technique that was recently applied to measure very accurately the ISF of Janus colloids in two dimensions~\cite{Kurzthaler2018}, and that of E. coli bacteria in 3D~\cite{Kurzthaler2024,Zhao2024}.

Starting from Eq. \eqref{eq:FPani}, Kurzthaler et al.~\cite{Kurzthaler2016} derived an explicit expression for the intermediate scattering function $F(\qq,t)$. Their strategy was to expand it on the functional basis associated with Eq. \eqref{eq:FPani}, which is a set of generalized spheroidal wave functions (for completeness, we recall their solution in Appendix \ref{app:Kurzthaler_result}). Their exact expression of $F(\qq,t)$  is therefore an infinite sum, whose terms involve functions and coefficients that are determined through the resolution of an eigenvalue problem. Importantly, the signature of activity and of the non-Gaussian behavior of the ABP is the emergence of oscillations at intermediate times in $F(q,t)$. We show on Fig. \ref{fig:schema_ell} the typical behavior of the ISF as a function of time, for a given value of the P\'eclet number  and for several values of the wavevector (see Section \ref{sec:simulations} and Appendix \ref{app:numerical} for details on the numerical simulations).

Since this resolution is only possible numerically in the general case, the solution is not fully analytical and remains rather implicit. In addition, this relies on the knowledge of the basis of functions associated with the Fokker-Planck equation obeyed by $P(\rr,\nn,t)$ [Eq. \eqref{eq:FPani}], i.e. one should be able to write the Fokker-Planck operator $\mathcal{L}$, which is such that $\partial_tP(\rr,\nn,t) = \mathcal{L}P$, in a basis where it is diagonal. There are more complex situations, for instance if the active particle experiences spatial confinement, if it is submitted to an external field, or if its orientation has non-Markovian dynamics, where these base functions may not be known, and may not allow such a direct expansion.
This calls for an alternative treatment of Eq. \eqref{eq:FPani}, that would be less specific and applicable to more general situations.

\subsection{Goals of the paper}

If the quantity of interest is the marginal probability distribution $\rho$, a direct way of treating Eq. \eqref{eq:FPani} is to naively integrate it over the orientation $\nn$. The issue is that, because of the translation-orientational coupling induced by the self-propulsion term, $\rho(\rr,t)$ does not obey a closed equation, but involves the polarization field $\boldsymbol{T}(\rr,t)$,
\begin{equation}
\label{eq:def_T}
\boldsymbol{T}(\rr,t)\equiv \int   \nn P(\rr,\nn;t) \dd{\nn} .
\end{equation}
In turn, the evolution equation for $\boldsymbol{T}(\rr,t)$ is not closed and involves the nematic tensor, defined as
\begin{equation}
\label{eq:def_Q}
\mathbf{Q}(\rr,t)\equiv \int\left(\nn\nn - \frac{1}{d} \mathbf{1}\right) P(\rr,\nn;t) \dd \nn ,
\end{equation}
where the part with the identity matrix ensures that the tensor $\mathbf{Q}$ is traceless. This expansion goes on: one could define higher-order moments that involve tensors of even higher ranks. One then faces an infinite hierarchy of equations, often called `moment expansion' in this context~\cite{Ahmadi2006, Marchetti2013, Saha2014, Solon2015a, Golestanian2022}. Formally, this can be seen as analogous to the BBGKY hierarchy~\cite{McQuarrie1976,Hansen2005}, which arises from pair interactions that couple degrees of freedom in multiparticle systems. In contrast, in the present case, the hierarchy equations stem from the coupling between positional and orientational degrees of freedom through self-propulsion.

A usual way to deal with this hierarchy is to adopt a truncation scheme, which is generally based on physical considerations. This greatly simplifies the set of equations and yields explicit (yet approximate) analytical results~\cite{Golestanian2012, Illien2017, Adeleke-Larodo2019, Adeleke-Larodo2019a, Golestanian2022}. However, very often, the range of validity of these closures is not studied thoroughly, in such a way that they appear as rather uncontrolled.
The goal of this paper is to systematically derive the equations obeyed by the moments $\rho$ (density), $\boldsymbol{T}$ (polarization), and $\mathbf{Q}$ (nematic tensor) for the general case of an anisotropic ABP. We then confront different closure schemes to numerical simulations and to the exact result from Ref.~\cite{Kurzthaler2016}, and we discuss their accuracy and their relevance to different situations of interest.

\section{Moment expansion}
\label{sec:FPE_momentexp_ani}

\subsection{Evolution equations of the moments}

We first note that the moments $\rho$, $\boldsymbol{T}$ and $\mathbf{Q}$ (defined in Eqs. \eqref{eq:def_rho}, \eqref{eq:def_T} and \eqref{eq:def_Q} respectively) appear naturally when performing an expansion of $P(\rr,\nn;t)$ in spherical harmonics:
\begin{equation}
    \label{eq:defplm}
    P(\rr,\nn;t) = \sum_{\ell=0}^{\infty} \sum_{m=-\ell}^{\ell} p_{\ell}^m(\rr,t) Y_\ell^{m}(\nn),
\end{equation}
with $p_\ell^m(\rr,t) = \int\dd\nn\,  P(\rr,\nn;t)Y_\ell^{m}(\nn)^*$ (the star $*$ denotes complex conjugation) and the $Y_\ell^{m}(\nn)$ are the spherical harmonics. The expansion of $P$ is therefore different from the one presented in Ref.~\cite{Kurzthaler2016}, which is performed in a more specific basis, namely generalized spheroidal wave functions.

The density field $\rho(\rr,t)$ can be expressed with the $p_{\ell=0}^0(\rr,t)$, the polarization field $\TT(\rr,t)$ with the $p_{\ell=1}^m(\rr;t)$ and the nematic field $\mathbf{Q}(\rr,t)$ with the $p_{\ell=2}^m(\rr;t)$, and so on. This is detailed in Appendix \ref{app:spherical_harmonics}. Thus, one can retrieve the evolution equations of the desired moment by setting a value for $\ell$ in the spherical harmonics expansion and by forming the right combination of coefficients $p_\ell^m(\rr,t)$. However, the general form of the expansion in spherical harmonics involves Clebsch-Gordan coefficients~\cite{Rose1957, Cohen-Tannoudji1977}, which makes this calculation quite cumbersome.

In practice, it is more convenient to multiply Eq. \eqref{eq:FPani} by  $1$, $n_i$ and $n_i n_j - \delta_{ij}/d$ respectively, and then integrate over all orientations. By doing so, one recovers the following set of equations, valid for $d=2$ and $d=3$, and where repeated indices implies summation,
\begin{align}
&\partial_t \rho(\rr,t) = D_0\nabla^2\rho - v\nabla\cdot\boldsymbol{T} + \Delta D \partial_j \partial_i Q_{ij} + \delta(\rr)\delta(t) ,\label{eq:dens_ani}\\
&\partial_t T_i(\rr,t) = D_1\nabla^2T_i - v\partial_j Q_{ij} - \frac{v}{d}\partial_i\rho - (d-1) D_r T_i + \Delta D \left( \frac{2}{d+2} \partial_i (\nabla\cdot\boldsymbol{T}) + \partial_j\partial_k K_{jki} \right), \label{eq:pola_ani}\\
&\partial_t Q_{ij}(\rr,t) = D_2\nabla^2Q_{ij} - v \partial_kK_{ijk} - 2dD_rQ_{ij}+\frac{2v 
}{d(d+2)} \delta_{ij}\nabla\cdot\boldsymbol{T} - \frac{v}{d+2} (\partial_i T_j + \partial_j T_i) \label{eq:nema_ani} \\
&+ \Delta D \left(\frac{2}{d(d+2)}\partial_j\partial_i\rho + \frac{2}{d+4}\partial_k (\partial_iQ_{jk} + \partial_jQ_{ik}) + \partial_k\partial_lM_{ijkl} - \frac{2}{d^2(d+2)}\delta_{ij}\nabla^2\rho - \frac{4}{d(d+4)}\delta_{ij}\partial_l\partial_kQ_{kl} \right) ,\nonumber
\end{align}
with 
\begin{eqnarray}
D_n &=& \frac{1}{d+2n}[(d+2n -1)D_{\perp} + D_{\parallel}], \\
\Delta D &=& D_{\parallel} - D_{\perp},
\end{eqnarray}
and where $K_{ijk}$ and $M_{ijkl}$ are the coefficients of the traceless tensors of rank 3 and rank 4 that appear in the moment expansion:
\begin{eqnarray}
    K_{ijk}(\rr,t) &=& \int\dd\nn \left[ n_i n_j n_k - \frac{1}{d+2}(\delta_{ij}n_k + \delta_{ik}n_j + \delta_{jk}n_i) \right]P(\rr,\nn;t), \\
    M_{ijkl}(\rr,t) &=& \int\dd\nn \biggl[ n_i n_j n_k n_l - \frac{1}{d+4}(\delta_{ij}n_k n_\ell + \delta_{jk} n_\ell n_i + \delta_{k\ell} n_i n_j + \delta_{\ell i}n_j n_k + \delta_{ik}n_j n_\ell + \delta_{j\ell}n_i n_k) \nonumber \\
    &&+ \frac{1}{(d+2)(d+4)}(\delta_{ij}\delta_{k\ell}+\delta_{jk}\delta_{\ell i}+\delta_{ik}\delta_j\ell) \biggr] P(\rr,\nn;t).
\end{eqnarray}
The last term in the right-hand side of Eq. \eqref{eq:dens_ani} encodes the initial condition: at $t=0$, the ABP is located at the origin, and its orientation is uniformly distributed. Eqs. \eqref{eq:dens_ani}-\eqref{eq:nema_ani} are central equations, and several comments follow: (i) they are exact, but each order involves moments of higher order; (ii) there are two reasons why the equations are not closed: for a given order $k$, the self-propulsion term, proportional to $v$, involves a term of order $k+1$, and the anisotropy, controlled by $\Delta D$, involves a term of order $k+2$; (iii) moments of even (resp. odd) order are even (resp. odd) with position $\rr$. (iv) In Eq. \eqref{eq:pola_ani} the position probability distribution $\rho(\rr,t)$ depends only on the divergence of the polarization field $\TT(\rr,t)$ and nematic tensor $\QQ(\rr,t)$. Therefore, in Fourier space, only the component parallel to the wavevector $\qq$ of $\tf{\TT}$ and $\tf{\QQ}$ is needed.

We now present the different closure schemes, and we give the analytical expressions of the ISF that result from these approximations. These expressions will be discussed and confronted to numerical simulations in Section \ref{sec:validity}.

\subsection{Closure at order 0}
\label{sec:closure0}

The closure at order 0, which consists in taking $\boldsymbol{T}(\rr,t)$ and all higher-rank tensors in the expansion equal to zero, makes the problem quite trivial. The equation obeyed by $\rho$ is simply that of a passive Brownian particle with diffusion coefficient $D_0=(2D_\perp+D_\parallel)/3$:
\begin{equation}
    \partial_t \rho (\rr,t) = D_0 \nabla^2 \rho,
\end{equation}
and the ISF is simply a decreasing exponential: $F_0 (\qq,t) = \ex{-D_0 q^2 t}$. We use the notation $F_0$ to indicate the order of closure used here.

\subsection{Finer closure at order 0}

As this previous case is not very relevant for an active system, one can look at a finer approximation, that at least accounts for the activity-induced enhancement of the diffusion coefficient at long times. To this end, one still assumes that the nematic tensor $\mathbf{Q}$ and the higher-order moments are zero, but the equation obeyed by $\TT$ [Eq. \eqref{eq:pola_ani}] is simplified by assuming that the system is considered over large times and large scales, so that $\partial_t \TT \simeq 0$ and $\nabla^2 \TT \simeq 0$.

Under these approximations, we have the following set of closed equations:
\begin{align}
    &\partial_t \rho (\rr,t) = D_0 \nabla^2 \rho - v \nabla \cdot \boldsymbol{T} + \delta(\rr)\delta(t), \label{eq:approx_05_rho}\\
    &\TT (\rr,t) = - \frac{v}{d(d-1)D_r} \nabla \rho,
    \label{eq:approx_05}
\end{align}
which leads to
\begin{equation}
    \partial_t \rho_0 (\rr,t) = D_{0,\text{eff}} \nabla^2 \rho_0 + \delta(\rr)\delta(t) ,
\end{equation}
with $D_{0,\text{eff}} = D_0 + \frac{v}{d(d-1)D_r}$.
Thus, at this level of closure, the distribution of position of the ABP is simply Gaussian but  with an activity-enhanced translational diffusion coefficient, which is characteristic of the ABP model. The associated ISF reads
\begin{equation}
    F_{0,\text{Act}}(q,t) = \ex{- D_{0,\text{eff}} q^2 t}.
    \label{eq:FGauss}
\end{equation}
Obviously, this order of approximation is not sufficient to account for the oscillations in the ISF, which can only be accounted for in a non-Gaussian theory.

\subsection{Closure at order 1}
\label{sec:closure1}

Closing one order higher, the terms $\partial_t\TT$ and $\nabla^2 \TT$ can be kept in Eq. \eqref{eq:pola_ani}, and one may only neglect moments of order $\mathbf{Q}$ and higher. Under these approximations, we have the following set of closed equations for $\rho$ and $\TT$:
\begin{align}
    \partial_t \rho (\rr,t) =& D_0 \nabla^2 \rho - v \nabla \cdot \boldsymbol{T} + \delta(\rr)\delta(t), \label{eq:rho_order1} \\
    \partial_t \TT(\rr,t) =& D_1\nabla^2 \TT - \frac{v}{d}\nabla \rho - (d-1) D_r \TT + \Delta D \frac{2}{d+2} \nabla (\nabla\cdot\boldsymbol{T}). \label{eq:T_order1}
\end{align}
These coupled equations can be solved in Fourier space for space and time. Throughout the paper, we will use the following convention for Fourier transformation for any function $\psi$:
\begin{equation}
    \widetilde\psi(\qq,\omega)=\int\dd\rr\int\dd t\,  \ex{-\ii\qq\cdot\rr} \ex{-\ii\omega t} \psi(\rr,t) \qquad ; \qquad 
    \psi(\rr,t) = \frac{1}{(2\pi)^{d+1}}\int\dd\qq\int\dd \omega\,  \ex{\ii\qq\cdot\rr} \ex{\ii\omega t} \widetilde\psi(\qq,t).
\end{equation}
Note that the tilde notation will also be employed when only spatial Fourier transform is performed: $\widetilde\psi(\qq,t) = \int\dd\rr\,  \ex{-\ii\qq\cdot\rr} \psi(\rr,t)$.
Taking the spatial and temporal Fourier transform of Eq. \eqref{eq:T_order1} leads to,
\begin{equation}
    \ii \omega  \widetilde{\boldsymbol{T}}(\qq;\omega)  = - D_1 q^2 \widetilde{\boldsymbol{T}} - \frac{\ii v}{d} \widetilde{\rho}  \qq - (d-1) D_r \widetilde{\boldsymbol{T}} - \Delta D \frac{2q^2}{d+2} \widetilde{\boldsymbol{T}}.
\end{equation}
Since the equation on the density only involves $\nabla \cdot \TT$, we project the equation for $\widetilde{\TT}$ on the wavevector $\qq$, and we get:
\begin{align}
    &\ii \omega \widetilde{\rho} (\qq,\omega) = - D_0 q^2 \widetilde{\rho} - \ii v (\boldsymbol{\widetilde{T}}\cdot \qq) + 1 \\
    &\ii \omega (\widetilde{\boldsymbol{T}}(\qq,\omega) \cdot \qq)  = - D_1 q^2 (\widetilde{\boldsymbol{T}} \cdot \qq) - \frac{\ii v q^2}{d} \widetilde{\rho} - (d-1) D_r (\widetilde{\boldsymbol{T}} \cdot \qq) - \Delta D \frac{2q^2}{d+2} (\widetilde{\boldsymbol{T}} \cdot \qq). 
\end{align}
Introducing ${T_{\parallel}}(\qq,\omega) \equiv {\boldsymbol{\widetilde{T}}\cdot\qq}/{\vert \qq \vert}$ and rearranging the terms: 
\begin{align}
    &(\ii \omega + D_0 q^2)\widetilde{\rho} (\qq,\omega) = -\ii v q {T_{\parallel}} + 1 \\
    &(\ii \omega + D_1 q^2 + \Delta D \frac{2q^2}{d+2} +(d-1)D_r) {T_{\parallel}}(\qq,\omega) = - \frac{\ii v q}{d} \widetilde{\rho}  \nonumber.
\end{align}
Finally, eliminating $T_\parallel$ yields
\begin{equation}
    \widetilde{\rho}(\qq,\omega) = \frac{\ii \omega + (D_1 + \frac{2}{d+2}\Delta D) q^2 + (d-1)D_r}{(\ii \omega + (D_1 + \frac{2}{d+2}\Delta D) q^2 + (d-1)D_r)(\ii \omega + D_0 q^2) + \frac{v^2q^2}{d}}.
\end{equation}
In order to obtain the ISF, we need to perform the inverse temporal Fourier transform, 
\begin{equation}
    F_1(\qq,t)=\int \frac{\dd \omega}{2\pi} \ex{\ii \omega t}  \frac{\ii \omega + D_1 q^2 + (d-1)D_r + \frac{2q^2}{d+2}\Delta D}{(\ii \omega + D_1 q^2 + (d-1)D_r + \frac{2q^2}{d+2}\Delta D)(\ii \omega + D_0 q^2) + \frac{v^2q^2}{d}} .
    \label{eq:before_residue}
\end{equation}
We then use the residue theorem. The poles of the denominator of the integrand in Eq. \eqref{eq:before_residue} read
\begin{equation}
\omega_{\pm} =
\begin{cases}
  \frac{1}{2}(\ii b \mp \sqrt{\delta}) & \text{if } \delta \geq 0, \\
  \frac{\ii}{2}(b \mp \sqrt{\vert \delta \vert}) & \text{if } \delta < 0.
\end{cases}
\end{equation}
with, 
\begin{align}
    b &= \left(D_1 + D_0 + \frac{2}{d+2}\Delta D\right)q^2 + (d-1)D_r, \\
    \delta &= - \frac{4(d-1)^2}{d^2(d+2)^2}\Delta D^2 q^4 + \frac{4}{d} \left( v^2 - \frac{(d-1)^2}{(d+2)}\Delta D D_r \right) q^2 - (d-1)^2 D_r^2 .
\end{align}
The signs of $\delta$ and $\omega_{\pm}$ yield three cases for the expressions of the ISF.
\begin{equation}
    F_1(q,t) =
\begin{cases}
    \left( \frac{b - 2D_0 q^2}{\sqrt{\delta}} \sin{\frac{\sqrt{\delta}}{2}t} + \cos{\frac{\sqrt{\delta}}{2}t} \right) \ex{-\frac{b}{2}t} & \text{if } \delta \geq 0, \\
    \left( \frac{b - 2D_0 q^2}{\sqrt{\vert\delta\vert}} \sinh{\frac{\sqrt{\vert\delta\vert}}{2}t} + \cosh{\frac{\sqrt{\vert\delta\vert}}{2}t} \right) \ex{-\frac{b}{2}t} & \text{if } \delta < 0  \text{ and } b \geq \sqrt{\vert\delta\vert}, \\
    \frac{1}{2\sqrt{\vert\delta\vert}} \left( 2D_0 q^2 - b + \sqrt{\vert\delta\vert} \right) \ex{-\frac{1}{2}(b + \sqrt{\vert\delta\vert})t} & \text{else.}
\end{cases}
\label{eq:F1}
\end{equation}
Importantly, at this level of approximation, the ISF clearly becomes non-Gaussian. We also get with the same procedure the expression of $T_{\parallel}(q,t)$,
\begin{equation}
    T_{\parallel}(q,t) =
\begin{cases}
    \frac{2\ii vq}{d\sqrt{\delta}} \sin{\frac{\sqrt{\delta}t}{2}} \ex{-\frac{b}{2}t} & \text{if } \Delta \geq 0, \\
    \frac{2\ii vq}{d\sqrt{\vert\delta \vert}} \sinh{\frac{\sqrt{\vert \delta \vert}t}{2}} \ex{-\frac{b}{2}t} & \text{if } \Delta < 0  \text{ and } b \geq \sqrt{\vert\Delta\vert}, \\
    -\frac{\ii vq}{d\sqrt{\vert\delta\vert}} \ex{-\frac{1}{2}(b + \sqrt{\vert\delta\vert})t} & \text{else.}
\end{cases}
\label{eq:T1}
\end{equation}

\subsection{Closure at order 2}

Finally, closing one order higher, we simply discard the terms involving the moments of order 3 and 4, namely $K_{ijk}$ and $M_{ijkl}$. Under these approximations we have the following set of closed equations: 
\begin{align}
    &\partial_t \rho_2 (\rr,t) = D_0 \nabla^2 \rho_2 - v \nabla \cdot \boldsymbol{T} + \Delta D \partial_i \partial_j Q_{ij} + \delta(\rr)\delta(t) \label{eq:order2_rho}\\
    &\partial_t T_i(\rr,t) = D_1\nabla^2T_i - v\partial_j Q_{ij} - \frac{v}{d}\partial_i\rho_2 - (d-1) D_r T_i + \frac{2}{d+2} \Delta D \partial_i (\nabla\cdot\boldsymbol{T})   \label{eq:order2_T}\\
    &\partial_t Q_{ij}(\rr,t) = D_2\nabla^2Q_{ij} - 2dD_rQ_{ij}+\frac{2v}{d(d+2)} \delta_{ij}\nabla\cdot\boldsymbol{T} - \frac{v}{d+2} (\partial_i T_j + \partial_j T_i) \nonumber \\
    &+ \Delta D \left(\frac{2}{d(d+2)}\partial_j\partial_i\rho_2 + \frac{2}{d+4}\partial_k (\partial_iQ_{jk} + \partial_jQ_{ik}) - \frac{2}{d^2(d+2)}\delta_{ij}\nabla^2\rho_2 - \frac{4}{d(d+4)}\delta_{ij}\partial_l\partial_kQ_{kl} \right) \label{eq:order2_Q}
\end{align}
In Fourier space, projecting Eq. \eqref{eq:order2_T} (resp. Eq. \eqref{eq:order2_Q}) on $\qq/q$ (resp $\qq\qq/q^2$), we get, with $Q_\parallel \equiv \widetilde{Q}_{ij} q_i q_j/q^2$:
\begin{align}
    &(\ii\omega + D_0q^2)\tf{\rho} + \ii vq {T}_{\parallel} + \Delta Dq^2{Q}_{\parallel} = 1 \label{eq:closure2_1}\\
    &\frac{1}{d}\ii vq\tf{\rho} + \left(\ii\omega + (D_1 + \frac{2}{d+2}\Delta D)q^2 + (d-1)D_r\right){T}_{\parallel} + \ii vq {Q}_{\parallel} = 0 \label{eq:closure2_2} \\
    &\frac{2(d-1)}{d^2 (d+2)}\Delta D q^2 \tf{\rho} + \frac{2(d-1)}{d(d+2)}\ii vq{T}_{\parallel} + \left(\ii\omega + (D_2 + \frac{4(d-1)}{d(d+4)}\Delta D)q^2 + 2dD_r\right){Q}_{\parallel} = 0 \label{eq:closure2_3}
\end{align}
This gives a linear set of equations for the unknowns $\tf{\rho}$, ${T}_{\parallel}$ and ${Q}_{\parallel} $, that can be solved explicitly. The resolution is given in Appendix \ref{app:solution_closure_order_2}. We eventually get:
\begin{equation}
    F_2(q,t) = \int \frac{\dd \omega}{2\pi} \widetilde{\rho}(q,\omega)\ex{\ii \omega t} = \sum_{i=1}^3 \prod_{j \neq i} \frac{\mathcal{P}_1(\omega_i)\ex{\ii \omega_i t}}{(\omega_i - \omega_j)}
    \label{eq:F2}
\end{equation}
\begin{equation}
    T_{\parallel,2}(q,t) = \int \frac{\dd \omega}{2\pi} T_{\parallel}(q,\omega)\ex{\ii \omega t} = \sum_{i=1}^3 \prod_{j \neq i} \frac{\mathcal{P}_2(\omega_i)\ex{\ii \omega_i t}}{(\omega_i - \omega_j)}
    \label{eq:T2}
\end{equation}
\begin{equation}
    Q_{\parallel,2}(q,t) = \int \frac{\dd \omega}{2\pi} Q_{\parallel}(q,\omega)\ex{\ii \omega t} = \sum_{i=1}^3 \prod_{j \neq i} \frac{\mathcal{P}_3(\omega_i)\ex{\ii \omega_i t}}{(\omega_i - \omega_j)}
    \label{eq:Q2}
\end{equation}
where $\omega_1$,  $\omega_2$ and  $\omega_3$ are the roots of the determinant of the linear system [Eqs. \eqref{eq:closure2_1}-\eqref{eq:closure2_3}], whose expressions are given in Appendix \ref{app:solution_closure_order_2}, and where
\begin{align}
    \mathcal{P}_1(\omega)    &= \omega^2 - \ii \left( \frac{92}{105} D_{\perp}q^2 \frac{118}{105}D_{\parallel}q^2 + 8 D_r \right)\omega  \nonumber \\
    &- \left( \frac{52}{105}D_{\parallel}D_{\perp}q^4 + \frac{11}{35}D_{\parallel}^2 q^4 + \frac{4}{21}D_{\perp}^2 q^4 + \frac{488}{105}D_{\parallel}D_rq^2 \frac{352}{105}D_{\perp}D_r q^2 + 12D_r^2 + \frac{4}{15}v^2 q^2 \right).
    \label{eq:P} \\
    \mathcal{P}_2(\omega) &= -\frac{vq}{3}\omega + \frac{\ii v q^3}{105} \left(26D_\perp + 9D_\parallel \right) + 2\ii vq D_r \label{eq:P2} \\
    \mathcal{P}_3(\omega) &= \frac{4\ii}{45}  (D_{\parallel}-D_\perp)q^2\omega  - \frac{4}{225} D_\parallel D_\perp q^4 - \frac{8}{225}D_\perp^2 q^4 + \frac{4}{75}D_\parallel^2 q^4 + \frac{8}{45}D_r (D_\parallel-D_\perp)q^2 + \frac{4}{45}v^2q^2 \label{eq:P3}
\end{align}
Together with Eq. \eqref{eq:F1}, Eqs. \eqref{eq:F2} and \eqref{eq:P} are one of the central equations of this article: they give a fully analytical expression for the ISF, that accounts for its non-Gaussian behaviour.

\subsection{Closure at higher orders}

As a remark, we note that, at the next order of closure, the denominator of the rational function that needs to be Fourier inverted will be a polynomial of degree 4 in $\omega$. Thus, there will not be an analytical expression of the ISF as there is no general expression for the roots of such polynomial. These poles can be easily found numerically, but then our strategy no longer has the advantage of being explicit. Therefore, we do not consider higher-order closures.

\section{Validity of the closures}
\label{sec:validity}

In this section, we perform Brownian dynamics simulations of a single anisotropic ABP in 3D. With the resulting trajectories, we compute the intermediate scattering function, polarization field, and nematic tensor in order to compare them to the analytical results of each closure and to study in which range of parameters the different closures are valid.

\subsection{Numerical simulations}
\label{sec:simulations}

To perform simulations, we start from the Langevin equations obeyed by $\rr$ and $\nn$:
\begin{eqnarray}
    \frac{\dd \rr}{\dd t}(t) &=& v \nn(t) + \sqrt{2 \mathbf{D}} \boldsymbol{\xi}(t), 
    \label{eq:Langevin_pos0} \\
    \frac{\dd \nn}{\dd t}(t) &=&  \sqrt{2 D_r}  \boldsymbol{\zeta}(t) \times  \nn(t),
    \label{eq:Langevin_ori0}
\end{eqnarray}
where $\boldsymbol{\xi}(t)$ and $\boldsymbol{\zeta}(t)$ are Gaussian white noises, i.e. $\langle \xi_i(t)\rangle = 0$ and $\langle \xi_i(t)\xi_j(t')\rangle = \delta_{ij}\delta (t-t') $ (and similarly for $\boldsymbol{\zeta}$). Importantly, in 3D, the noise in Eq. \eqref{eq:Langevin_ori0} is multiplicative, and it must be interpreted in the Stratonovitch way, to ensure that the norm of $\nn$ remains unity.

Integrating Eq. \eqref{eq:Langevin_pos0} and Eq. \eqref{eq:Langevin_ori0} over a timestep $\Delta t$, we get:
\begin{eqnarray}
    r_i(t+\Delta t) &=& r_i(t) + v n_i(t) \Delta t + \left[ \sqrt{2D_{\parallel}} n_i(t) n_j(t) + \sqrt{2D_{\perp}} (\mathbf{1} - n_i(t) n_j(t)) \right] \mathcal{N}(0,1) \sqrt{\Delta t}, \\
    n_i(t+\Delta t) &=& n_i(t) - 2  D_r n_i(t) \Delta t - \sqrt{2D_r} \epsilon_{ijk} n_j(t) \mathcal{N}(0,1) \sqrt{\Delta t},
    \label{eq:ni_discretized}
\end{eqnarray}
where $\mathcal{N}(0,1)$ are random numbers drawn from a normal distribution of average 0 and unit variance.
The second term in Eq. (\ref{eq:ni_discretized}) is a consequence of the Stratonovitch interpretation of Eq. \eqref{eq:Langevin_ori0} when integrating it over a timestep $\Delta t$. This point is rather subtle, and was discussed recently by other authors~\cite{Kurzthaler2016, Höfling2025}. For completeness, we give details on the treatment of Eq. \eqref{eq:Langevin_ori0} in Appendix \ref{app:multiplicativenoise}.

This model has four parameters: the three diffusion coefficients $D_{\parallel}$, $D_{\perp}$ and $D_r$, and the self-propulsion velocity $v$. Although the results hold for a generic anisotropic particle, we will consider the case of a prolate ellipsoid in order to relate anisotropy in diffusion coefficients to anisotropy in aspect ratio. This is done using Perrin formulae, which gives the translational and rotational friction coefficients of ellipsoidal objects as a function of their major axis $a$, minor axis $b$ and the viscosity of the medium (see Appendix \ref{app:numerical}).

Following Ref.~\cite{Kurzthaler2016}, we define two dimensionless parameters. First, we introduce the translational anisotropy relative to the mean diffusion coefficient:
\begin{equation}
    \frac{\Delta D}{D_0} = \frac{(D_\parallel-D_\perp)}{(D_\parallel+2D_\perp)/3}
\end{equation}
In the following, this parameter will be set to $\Delta D / D_0 = 0.21$, which corresponds to a prolate ellipsoid of aspect ratio $a/b = 3$. The influence of this parameter on the signature of the activity is discussed in Appendix \ref{app:aspect_ratio}.

The second parameter is the P\'eclet number $\Pe$, which measures the importance of the active motion relative to translational diffusion. We take as a characteristic length of the system $\ell_0=\sqrt{D_0/D_r}/2$, which corresponds to the radius of a spherical particle that experiences translational and rotational diffusion with coefficients $D_0$ and $D_r$. The P\'eclet number is then defined by, 
\begin{equation}
    \Pe = \frac{v\ell_0}{D_0}
\end{equation}
This is our main parameter of interest. In practice, it will be tuned by changing the self-propulsion velocity. In our simulations, the P\'eclet number will typically vary from 5 to 40, which is relevant to describe a broad range of real active particles~\cite{Bechinger2016}. For instance, artificial colloidal active systems such as Janus particles,  typically have a P\'eclet number around $\Pe \sim 5$~\cite{Bechinger2016, Kurzthaler2018}. Swimming bacterial cells, on the other hand, generally have higher self-propulsion velocity and therefore have a P\'eclet number around $\Pe \sim 20$~\cite{Kurzthaler2024}.

Finally, we make time and distances dimensionless with the typical time it takes to the particle to re-orient $\tau_{\mathrm{rot}} = 1/D_r$ and the persistence length $L_p = v/D_r$.

\subsection{A first insight into the ISF}

\begin{figure}
\begin{center}
\includegraphics[width= \textwidth]{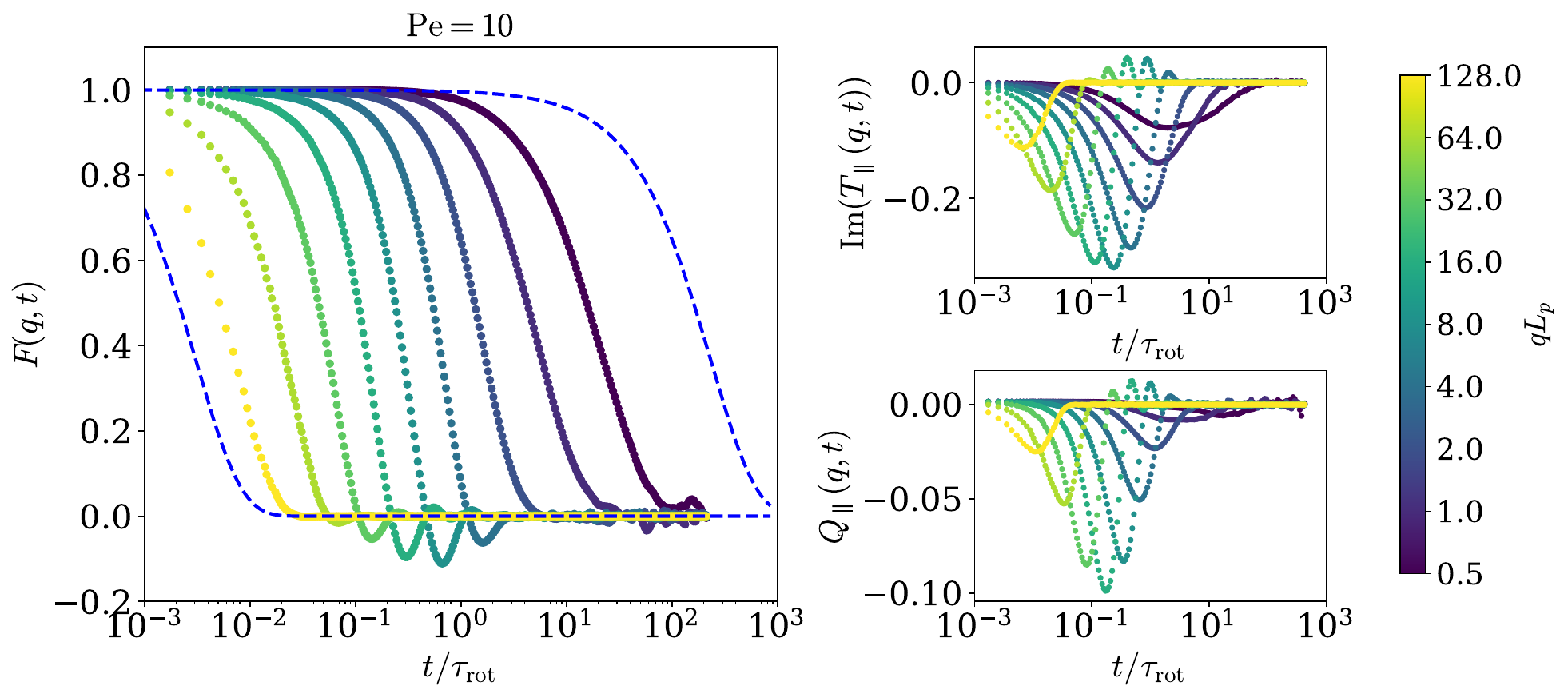}
\caption{{Left:} Simulated intermediate scattering function $F(q,t)$ of an anisotropic ABP as a function of time and for different wavenumbers $q$. The dashed blue lines represents the effective diffusion regimes that are retrieved at large and small wavenumbers. {Right:} Spatial Fourier transform of the first and second orientational moments (top: polarization; bottom: nematic tensor) projected along the wavenumber $q$. Parameters: $\Delta D/D_0 = 0.21$, $a/b = 3$, $L_p = 14$, $\tau_{\text{rot}} = 1.17$.}
\label{fig:isf_tpar_qpar}
\end{center}
\end{figure}

The left panel of Fig.~\ref{fig:isf_tpar_qpar} shows the ISF $F(q,t)$ as a function of time for different wavenumbers $q$, as measured in numerical simulations. The wavenumber $q$ can be interpreted as the scale at which the system is observed. For large $q$, the system is observed from very close. At this scale, the ABP has a purely diffusive behavior. Indeed,  at very small scales, the activity is indiscernible from translational diffusion, and the ISF is $F(q,t) \simeq \ex{-D_0 q^2 t}$. In the opposite limit of very small $q$, i.e. when the  system is observed from very far, the ABP behaves as a Brownian particle with effective diffusion coefficient $D_{0,\text{eff}} = D_0 + \frac{v}{d(d-1)D_r}$, and the ISF is approximately $F(q,t) \simeq \ex{-D_{0,\text{eff}} q^2 t}$. Consequently, in these two extreme regimes of wavevectors (which are shown as dashed lines on Fig.~\ref{fig:isf_tpar_qpar}(left)), the ISF simply has  a decreasing exponential behavior. 

For intermediate values of $q$, the Gaussian behavior is also retrieved at small and large times. The behavior of the ISF becomes more complicated at intermediate times, where we can observe oscillations that are the signature of activity.
This signature is what we want to characterize with our analytical results as it corresponds to the deviation from the Gaussianity.

As discussed in Section \ref{sec:FPE_momentexp_ani}, because of anisotropy and self-propulsion, the ISF is related to the first and second moments of the orientational distribution $T_\parallel(q,t)$ and $Q_\parallel(q,t)$ projected along the wavenumber $q$. We plot these quantities  on the right of Fig. \ref{fig:isf_tpar_qpar}. We can observe that these two terms are generally non-zero.

Beyond the intermediate scattering function, our framework naturally provides access to the polarization and nematic fields, quantities that are rarely measured in experiments. Extending techniques such as differential dynamic microscopy to probe not only the positional dynamics but also orientational moments would open new perspectives. Direct measurements of polarization or nematic correlations would offer valuable insight into the coupling between translation and orientation, and would allow a more stringent test of theoretical predictions based on moment expansions.

\subsection{Confronting simulations and analytical results}

\subsubsection{Validation of hypotheses at the finer closure at order 0}

\begin{figure}
\begin{center}
\includegraphics[width= 0.95\textwidth]{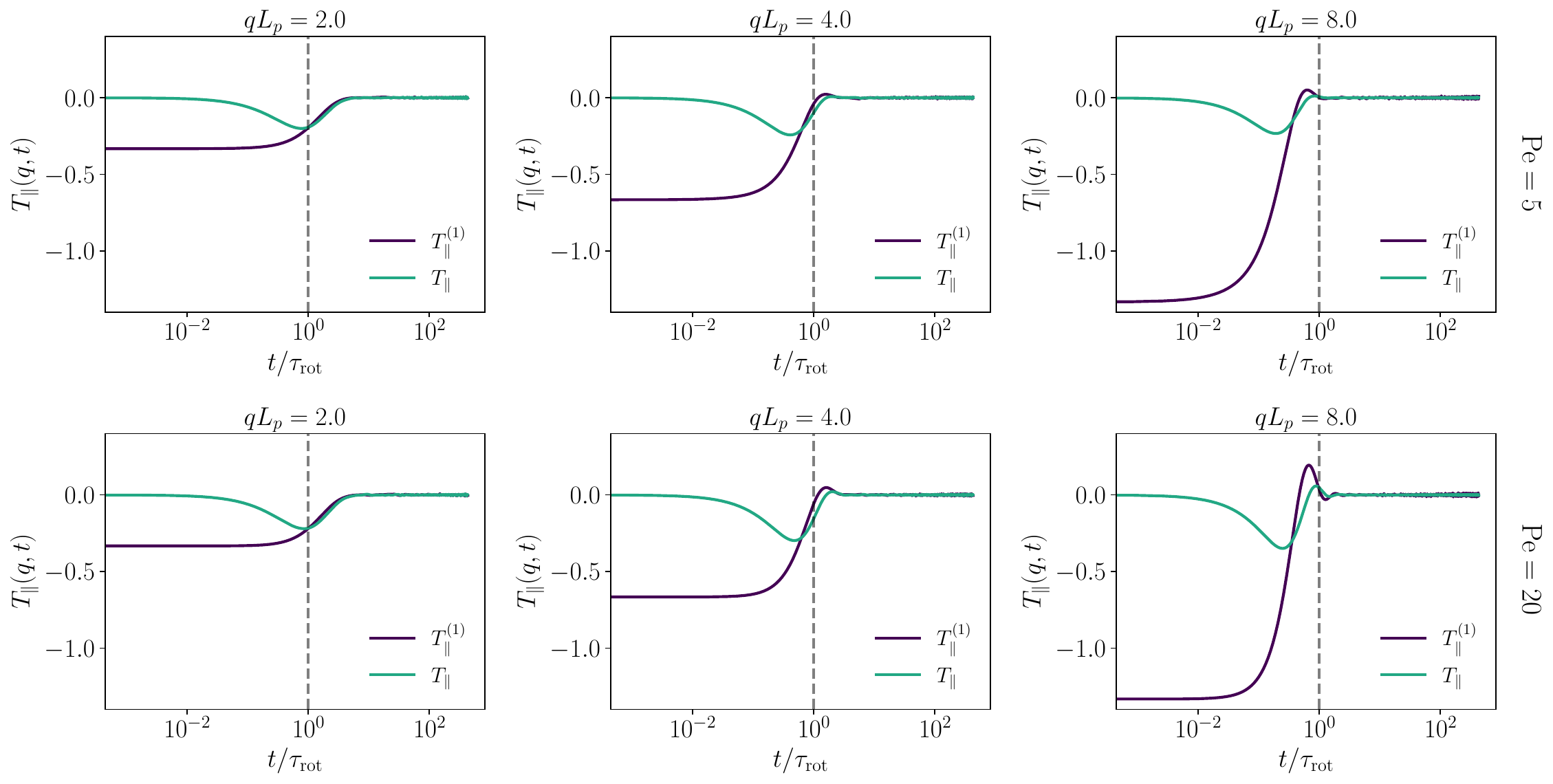}
\caption{Comparison between the imaginary part of the spatial Fourier transform of the polarization $T_{\parallel}(q,t)$ and the imaginary part of the spatial Fourier transform of the gradient of the ISF for different wavenumbers $q$. Parameters:  $\Delta D/D_0 = 0.21$, $a/b = 3$, $L_p = [7.0,28.1]$ (for the respective P\'eclet numbers), $\tau_{\text{rot}} = 1.17$.}
\label{fig:compa_tpar_qisf}
\end{center}
\end{figure}

In many active matter models \cite{Golestanian2012, Golestanian2022,Illien2017}, assuming the polarization field $\TT$ to be proportional to the gradient of the density $\rho$ [Eq. \eqref{eq:approx_05}] is a simple approximation that leads to the finer closure at order 0. It relies on the following hypotheses:
\begin{enumerate}
    \item[\textbf{(1)}] $\partial_t \boldsymbol{T}(\rr,t) \simeq 0$ (i.e. $\boldsymbol{T}(\rr,t)$ is stationary),
    \item[\textbf{(2)}] $\nabla^2 \boldsymbol{T}(\rr,t) \simeq 0$,
    \item[\textbf{(3)}] $\partial_j Q_{ij} \simeq 0$,
    \item[\textbf{(4)}] Higher-rank moments are all vanishing.
\end{enumerate}
These approximations lead to Eq. \eqref{eq:approx_05}, which in Fourier space gives, 
\begin{equation}
    {T}_{\parallel}^{(1)}(q,t) = - \ii\frac{v}{6D_r}q F(q,t)
    \label{eq:H0eq}
\end{equation}
The comparison between the polarization $T_\parallel$ measured in numerical simulations and its approximate expression ${T}_{\parallel}^{(1)}$ is shown in Fig. \ref{fig:compa_tpar_qisf}. First, as expected from hypothesis \textbf{(1)}, this approximation is only valid for large enough time (typically $t \gtrsim 1/D_r$). Second, as the wavenumber increases, ${T}_{\parallel}^{(1)}$ deviates from $T_\parallel$: this is consistent with the fact that hypothesis \textbf{(2)} only holds at large enough distances. Finally, for very high P\'eclet numbers, the approximation fails: indeed, higher orders of the moment expansion should be kept to make accurate predictions when activity increases (hypotheses \textbf{(3)} and \textbf{(4)}).

\subsubsection{Comparison between closures at order 1 and 2}

\begin{figure}
\begin{center}
\includegraphics[width= 0.8\textwidth]{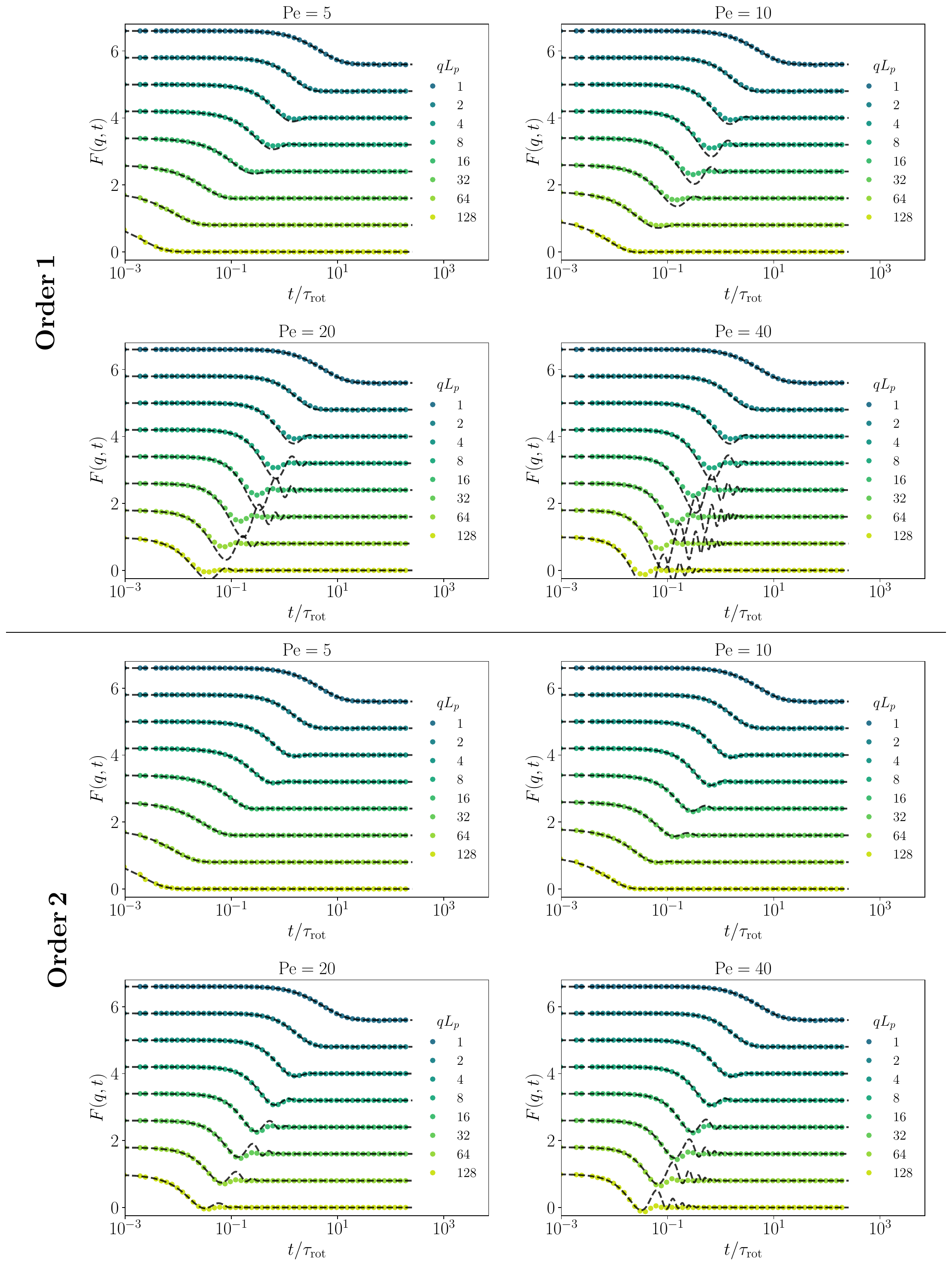}

\caption{Comparison between the ISF measured in numerical simulations (colored symbols) and the analytical ISF computed with the closures at order 1 and 2 (blacked dashed line, Eqs. \eqref{eq:F1} and \eqref{eq:F2} respectively)  for different wavenumbers $q$ and different P\'eclet number $\Pe$. The different plots are shifted in the vertical direction for clarity -- all the ISF actually tend to $1$ in the limit $t\to 0$ and $0$ in the limit $t\to\infty$. Parameters:  $\Delta D/D_0 = 0.21$, $a/b = 3$, $L_p = [7.0,14.0,28.1,56.2]$ (for the respective P\'eclet numbers), $\tau_{\text{rot}} = 1.17$.}
\label{fig:compa_theo_simu1}
\end{center}
\end{figure}

We show on Fig. \ref{fig:compa_theo_simu1} the comparison between the  ISF measured from numerical simulations, and our analytical results with closures at order 1 and 2, i.e. with Eqs. \eqref{eq:F1} and \eqref{eq:F2} respectively, for various P\'eclet numbers and wavenumbers.
The closure schemes we put forward provide reasonable estimates of the ISF, provided that the P\'eclet number is not too large, and that we consider wavevectors that are close enough to the lower bound $q \sim 1/L_p$. When the hierarchy of equations is closed at order $1$, the oscillatory signature is captured, but the amplitude of the oscillations are much larger than expected when the wavenumber increases. Finally, at order 2, this mismatch to the exact solution is smaller, and the closure holds for larger wavenumbers $q$.  

\begin{figure}
\begin{center}
\includegraphics[width= 0.5\textwidth]{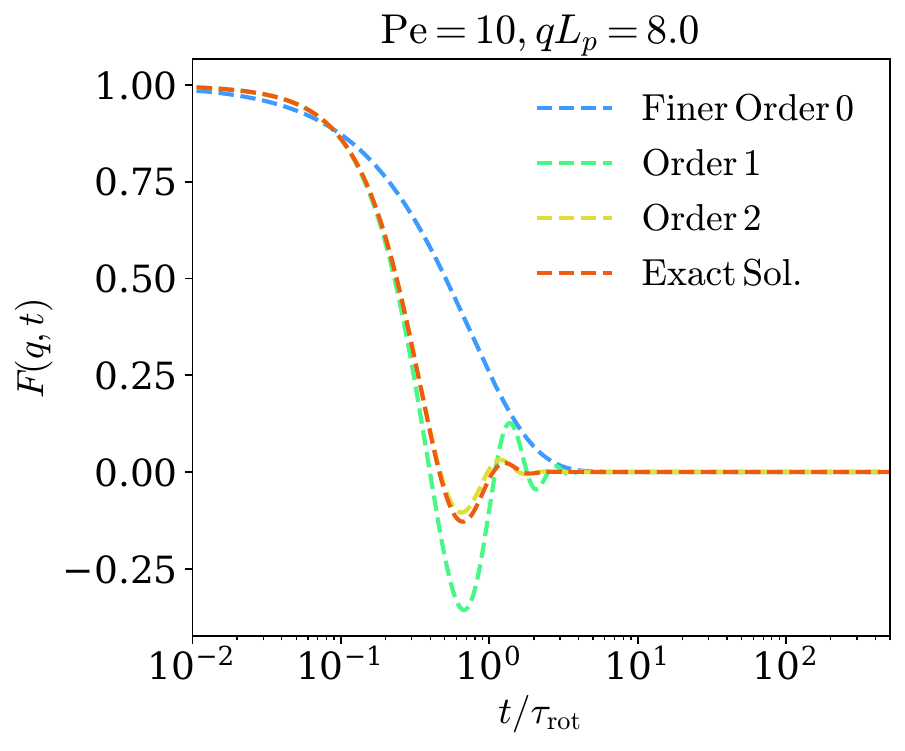}
\caption{Comparison between the analytical expressions obtained with the different closures and the exact solution as a function of time. Parameters: $\Delta D/D_0 = 0.21$, $a/b = 3$, $L_p = 14$, $\tau_{\text{rot}} = 1.17$.}
\label{fig:compa_theo_simu2}
\end{center}
\end{figure}

One can also notice that there is a considerable improvement between the different closures: this can be seen on Fig. \ref{fig:compa_theo_simu2}, where we compare the finer closure at order $0$, the closures at order $1$ and $2$, and the exact solution, for given values of the P\'eclet number and the wavevector. At order $0$, the signature of activity is of course not captured at all, as it predicts an effective purely diffusive regime, which leads to $F_{0,\text{Act}}(q,t)=\ex{-D_{0,\text{eff}}q^2t}$. 

\subsubsection{Error relative to the exact solution}

\begin{figure}
\begin{center}
\includegraphics[width= 0.55\textwidth]{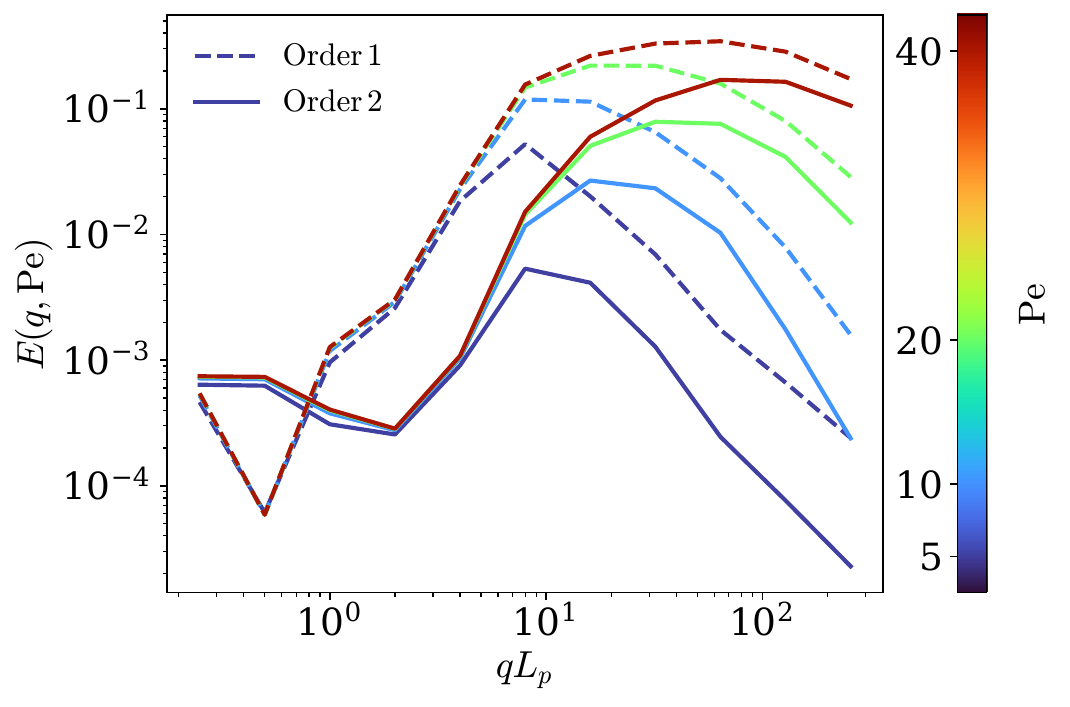}
\caption{Error between the ISF estimated from closures at orders 1 and 2 and the exact solution, as defined in Eq. \eqref{eq:error}, and as a function of the rescaled wavevector $q L_p$, for different values of $\Pe$. Parameters:  $\Delta D/D_0 = 0.21$, $a/b = 3$, $L_p = [7.0,14.0,28.1,56.2]$ (for the respective P\'eclet numbers), $\tau_{\text{rot}} = 1.17$.}
\label{fig:error}
\end{center}
\end{figure}

In order to quantify the validity of our approximations, we define the error $E(q,\Pe)$ as:
\begin{equation}
    E(q,\Pe) =  \langle |F_i(q,t)-F_\text{exact}(q,t)|  \rangle_t
    \label{eq:error}
\end{equation}
for $i=1,2$, and where the average $\langle \cdot \rangle_t$ runs over a two-decade-long window, centered on the time $t_{1/2}$, which is such that $F(q,t=t_{1/2})=1/2.$ This error function, which is plotted on Fig. \ref{fig:error}, quantifies what is observed on Fig. \ref{fig:compa_theo_simu1}: (i) Order 2 is almost always more accurate than order 1; (ii) Both approximations are equally good for very small wavevectors -- this is expected since the closures typically rely on large-distance approximations. However, order 2 provides very good approximation for intermediate wavevectors, up to $qL_p \simeq 10$; (iii) Finally, for large wavevectors, the approximations only hold for small P\'eclet numbers. 

Finally, we also explored how the aspect ratio of the ABP (i.e. the ratio $\Delta D/D_0$) affects the validity of the closure approximations at a constant P\'eclet number (i.e. by changing the self-propulsion velocity accordingly). The amplitude of the activity signature is affected by the anisotropy, but its period is unchanged. Interestingly, anisotropy does not appear to influence the validity of the presented closures, as the error behaves similarly (see Appendix \ref{app:aspect_ratio}). This validates our approximate, yet explicit, expression of the ISF for a wide range of aspect ratio.

\section{Discussion and perspectives}

A central outcome of our study is that the choice of closure in the moment expansion is not universal but depends on the spatio-temporal scale at which the active Brownian particle is probed. At very large scales and long times, the order-0 finer closure already provides a faithful description: the system effectively behaves as a passive Brownian particle with an activity-enhanced diffusion coefficient. At intermediate scales, however, non-Gaussian features, such as oscillations in the ISF, become essential. These signatures are captured only once polarization and nematic contributions are retained, i.e. at closures of order 1 and 2. The closure at order 2 gives clearly the best approximate expression, but the expression at order 1 has the benefit of being very explicit and more physical. 

Our comparison with simulations provides a practical guide. At small wavevectors (i.e. large distances), the finer closure at order 0 [Eq. \eqref{eq:FGauss}] is sufficient. At moderate wavevectors and moderate Péclet numbers, order 1 closure [Eq. \eqref{eq:F1}] captures the onset of activity-induced oscillations. At intermediate wavevectors and/or higher Péclet numbers, order-2 closure [Eq. \eqref{eq:F2}] is necessary to achieve quantitative agreement. Beyond this regime, higher-order truncations are formally possible but lose their main advantage, i.e. being explicit.

This analysis highlights the practicality of the moment expansion: despite its approximate nature, it yields closed-form results that can be directly compared to experimental observables. Moreover, the framework is flexible and modular, which makes it particularly attractive for more complex extensions. For instance, one could consider other dynamics for activity, such as run-and-tumble or run-and-reverse with stochastic residence times, whether exponentially distributed (Markovian) or with more general, non-Markovian waiting-time statistics. Likewise, external fields or confinement can be introduced by adding corresponding terms in the Fokker–Planck operator, and the moment expansion remains applicable without the need to identify a specific functional basis for exact diagonalization.

In this sense, our approach provides a general method that complements existing exact formulations. While the latter are elegant and mathematically complete, they typically rely on bases (such as spheroidal wave functions) that may be unknown or impractical in more involved situations. Moment closures, on the other hand, offer a systematic and explicit way forward, serving as a practical guide for both theorists and experimentalists to select the level of approximation best suited to the spatiotemporal regime of interest.

\section*{Acknowledgements}

T. G. acknowledges funding from Institut des Sciences du Calcul et des Donn\'ees (Sorbonne Université). P. I. thanks Ramin Golestanian and Fanlong Meng for discussions during early stages of this work. We thank Christina Kurzthaler for sharing with us a Mathematica notebook that implements the resolution detailed in Appendix \ref{app:Kurzthaler_result}, and that we used for the plots shown on Fig. \ref{fig:schema_ell}.

\appendix

\section{Numerical simulations}
\label{app:numerical}

\subsubsection{Presentation of the simulations}

Simulations are performed in 3D without periodic boundary conditions (in order to avoid restrictions on the values of the wavenumbers). In order to simulate the ISF, we use a sliding average and also average over many simulations with the same parameters but with different seeds. Formally, we compute the ISF doing,
\begin{equation}
    F(q,m\Delta t) = \frac{1}{N_s} \sum_{s = 1}^{Ns} \left[ \frac{1}{N_t - m} \sum_{k=0}^{N_t -m-1} \ex{- \ii \qq \cdot [\rr_s((m+k)\Delta t) - \rr_s(k\Delta t)]} \right]
\end{equation}
with, $N_t$ the number of simulation steps, $\Delta t$ the simulation timestep and $N_s$ the number of different simulations. This computation can be seen as the computation of an autocorrelation function, and therefore we can use a FCA (Fast Correlation Algorithm) in order to improve the analysis time of the trajectories~\cite{kneller1995nmoldyn}. $T_{\parallel}(q,t)$ and $Q_{\parallel}(q,t)$ are computed similarly using a FCA method.

\subsubsection{Parameters choices}

To take into account the ellipsoidal shape, we use the generalization of Stokes's law to ellipsoids, called the Perrin factors~\cite{Perrin1934,Perrin1936}. These factors come from the derivation of the flow around an ellipsoidal geometry and are given by~\cite{theers2018clustering}
\begin{align}
    \gamma_{t,\parallel} & = \frac{8}{3} \gamma_{t,\text{iso}}(-2e + (1+e^2)L)^{-1} \\
    \gamma_{t, \perp} &=  \frac{16}{3}\gamma_{t,\text{iso}}e^3(2e + (3e^2 - 1)L)^{-1} \\
    \gamma_{r, \parallel} &=\frac{4}{3} \gamma_{r,\text{iso}} e^3(1-e^2)(2e - (1 - e^2)L)^{-1} \\
    \gamma_{r, \perp} &=  \frac{4}{3}\gamma_{r,\text{iso}} e^3(2-e^2)(-2e + (1 + e^2)L)^{-1}
\end{align}
with
\begin{itemize}
    \item $\gamma_{t,\text{iso}} = 6\pi\eta a/2$ : translational friction coefficient of a sphere of diameter $d_x$.
    \item $\gamma_{r,\text{iso}} = 8\pi\eta(a/2)^3$ : rotational friction coefficient of a sphere of diameter $d_x$
    \item $\eta$ : viscosity of the medium

    \item $e = \sqrt{1-(b/a)^2}$ :  eccentricity 
    \item $L = \log{\frac{1+e}{1-e}}$
\end{itemize}

This allows us to reduce the number of independent parameters in the system. With this definition, which encapsulates the hydrodynamics around an ellipsoid at a low Reynolds number, friction depends only on the aspect ratio $a/b$ of the particle and the viscosity $\eta$ of the medium. Also, in our case, we assume that the ellipsoid does not rotate around its main axis and, therefore, $\gamma_{r,\parallel}=0$.

We also recall here that the friction coefficients are linked to the diffusion coefficients through the Stokes-Einstein relation,
\begin{equation}
    D = \frac{k_BT}{\gamma}
\end{equation}
Thus, in the ellipsoidal case we have, 
\begin{equation}
    D_{t,\parallel} = \frac{k_BT}{\gamma_{t,\parallel}} , \quad D_{t,\perp} = \frac{k_BT}{\gamma_{t,\perp}} \quad \text{and} \quad D_r = \frac{k_BT}{\gamma_r}.
\end{equation}

Simulations are made in LJ units as it is more convenient to work with unitary parameters. This implies a choice in the parameters that are used to make our physical quantities dimensionless. As a unit length, we choose the smallest diameter of the ellipsoid $b$ and set it to 1 in simulation units. As we are in the overdamped limit, it is not relevant to define a unit mass, instead we take the thermal energy $k_B T$ as a unit of energy and set it to 1. Finally, we set $6\pi\eta=1$ in the definition of the friction coefficients. This means that the translational diffusion coefficient of a spherical particle of diameter $b$ is equal to 1.

\section{Reminder: Solution in generalized spheroidal wave functions}
\label{app:Kurzthaler_result}

In this Appendix, we recall for completeness the results from Ref.~\cite{Kurzthaler2016}, in which the authors give an expression of the intermediate scattering function (ISF) $F(\qq,t)$. They find the exact solution:
\begin{equation}
\label{eq:exact}
F(\qq,t) = \frac{1}{2} \ex{-D_\perp k^2 t} \sum_{\ell = 0}^\infty \ex{-D_r A_\ell^0 t } \left[ \int_{-1}^1 \dd\eta \, \text{Ps}_{\ell}^0(c,R,\eta)\right]^2,
\end{equation}
where the generalized spheroidal wave functions $\text{Ps}_{\ell}^m$ solve the eigenvalue problem,
\begin{equation}
\label{eq:eigenvaluepb}
\left[ \frac{\dd}{\dd \eta} \left( (1-\eta^2)\frac{\dd}{\dd \eta} \right) +R\eta -c^2\eta^2-\frac{m^2}{1-\eta^2} + A_\ell^m\right]\text{Ps}_{\ell}^m(c,R,\eta)=0
\end{equation}
where the dimensionless parameters $R=-\ii k v/D_r$, $c^2 = (D_{\parallel}-D_{\perp}) k^2/D_r$ and $\eta = \cos(\theta)$ have been defined.
Note that the term "generalized" is used as the functions $\text{Ps}_\ell^m$ are a generalization of the prolate spheroidal wave functions. Thus, when taking $R=0$ (i.e. $v=0$, the case of a passive Brownian particle), the eigenvalue problem defining the prolate spheroidal wave functions is retrieved. When one also takes $c=0$ (i.e. $\Delta D = 0$, the case of an isotropic passive Brownian particle) the eigenvalue problem defining the spherical harmonics $Y_\ell^m$ is retrieved.
In order to numerically evaluate the analytical solution recalled in Eq. \eqref{eq:exact}, the eigenvalues $A_\ell^0$ and the integrals over the eigenfunction $\text{Ps}_\ell^0$ are needed. To do so, the generalized spheroidal wave functions are expanded in terms of the Legendre polynomials.
\begin{equation}
    \text{Ps}_\ell^0(c,R,\eta) = \sum_{j=0}^{\infty} d_j^{0\ell} \vert j \rangle
\end{equation}
with, 
\begin{equation}
    \vert j \rangle = \sqrt{(2\ell+1)/2} P_j(\eta)
\end{equation}
where the $P_j(\eta)$ are the Legendre polynomials. \par
By inserting the expansion in eq.(\ref{eq:exact}) and eq.(\ref{eq:eigenvaluepb}) one gets,
\begin{equation}
    \label{eq:isf_expanded}
    F(q,t) = \ex{-D_{\perp}q^2t}\sum_{\ell=0}^\infty [d_0^{0\ell}]^2 \ex{-D_rA_\ell^0 t}
\end{equation}
and the following eigenvalue problem,
\begin{equation}
    \mathbf{B}(c,R,\eta)\mathbf{d}^{0\ell}(c,R) = A_\ell^0(c,R)  \mathbf{d}^{0\ell}(c,R)
\end{equation}
where, 
\begin{align}
    &\mathbf{d}^{0\ell} = (d_0^{0\ell}, d_1^{0\ell}, d_2^{0\ell},...) \\
    &(\mathbf{B}(c,R,\eta))_{jn} = \langle n \vert c^2\eta^2 - R\eta \vert j\rangle + n(n+1)\delta_{jn} \nonumber \\
    & \langle n \vert \eta^k \vert j \rangle = \sqrt{(2n+1)(2j+1)}/2 \int_{-1}^1 \dd \eta P_n(\eta)\eta^k P_j(\eta) \nonumber
\end{align}
In order to retrieve the eigenvectors $\mathbf{d}^{0\ell}$ and eigenvalues $A_\ell^0$, one needs to numerically evaluate the eigenvalue problem since $\mathbf{B}(c,R,\eta)$ is an infinite matrix. In~\cite{Kurzthaler2016}, the matrix is truncated at a sufficiently high order such that the normalization at time $t=0$ for the ISF is achieved. The truncation of the matrix also corresponds to a truncation of the sum in Eq. \eqref{eq:isf_expanded}. Note that the matrix $\mathbf{B}(c,R,\eta)$ is in fact a band matrix where the elements are non-vanishing only for $j=n-2,...,n+2$. This can be seen by evaluating $\langle n \vert \eta^k\vert j \rangle$ for $j=1,2$ with the following recursive relation on the Legendre polynomials, 
\begin{equation}
    \eta P_n(\eta) = \frac{1}{2n+1}((n+1)P_{n+1}(\eta) + n P_{n-1}(\eta)).
\end{equation}

\section{Expansion on spherical harmonics}
\label{app:spherical_harmonics}

In this Appendix, we show that expanding the Fokker-Planck equation [Eq. \eqref{eq:FPani}] in terms of the moments $\rho$, $\TT$, $\mathbf{Q}$ (and so on), is equivalent to an expansion on spherical harmonics. To the best of our knowledge, this relationship has not been clarified before.
The expansion  of $P(\rr,\nn;t)$ on spherical harmonics reads:
\begin{equation}
    \label{eq:defplm}
    P(\rr,\nn;t) = \sum_{\ell=0}^{\infty} \sum_{m=-\ell}^{\ell} p_{\ell}^m(\rr,t) Y_\ell^{m}(\nn)
\end{equation}
with,
\begin{equation}
    p_\ell^m(\rr,t) = \int\dd\nn \, P(\rr,\nn;t)Y_\ell^{m}(\nn)^*
\end{equation}
where the $Y_\ell^{m}(\nn)$ are the spherical harmonics in the usual convention from quantum mechanics~\cite{Rose1957, Cohen-Tannoudji1977}, and where $^*$ denotes complex conjugate. In the following, double sums on $\ell\geq 0$ and $-\ell\leq m \leq \ell$ will be denoted $\sum_{\ell,m}$. 

For the sake of clarity, the following calculation will be given for the case of an isotropic ABP (i.e. $D_\perp = D_\parallel$), but it is straightforward to extend it to the more general case of an anisotropic ABP. By multiplying Eq. \eqref{eq:FPani} by $Y_\ell^{m}(\nn)^*$ and integrating over all orientations, one can obtain the equation verified by the coefficients $p_\ell^m(\rr,t)$: 
\begin{equation}
    \partial_tp_\ell^m(\rr,t) = D\nabla^2p_\ell^m(\rr,t) -D_r \int \dd\nn Y_\ell^{m}(\nn)^*\rotop^2P(\rr,\nn;t) -v\nabla\cdot \int \dd\nn Y_\ell^{m}(\nn)^* \nn P(\rr,\nn;t)
\end{equation}
By making 2 integrations by part on the second term of the right-hand side of the equation (constant terms are zero because of periodicity), we find:
\begin{equation}
    \int \dd\nn Y_\ell^{m}(\nn)^*\rotop^2P(\rr,\nn;t) = \int \dd\nn P(\rr,\nn;t)\rotop^2Y_\ell^{m}(\nn)^*
\end{equation}
We then use that spherical harmonics are eigenfunctions of the rotational gradient operator with eigenvalues:
\[ \lambda_d(\ell) = \begin{cases} 
      \ell(\ell+1) & \text{if} \quad d=3, \\
      -\ell^2 & \text{if} \quad d=2,
   \end{cases}
\]
and we denote by $\mathbf{c}_\ell^m(\rr,t)$ the coefficients of the expansion on spherical harmonics of the product $\nn P(\rr,\nn;t)$. Furthermore, using the orthogonality of spherical harmonics, we find the equation for the coefficients $p_\ell^m$:
\begin{equation}
    \partial_tp_\ell^m(\rr,t) = D\nabla^2p_\ell^m(\rr,t) -\lambda_d(\ell)D_rp_\ell^m(\rr,t) -v\nabla\cdot\mathbf{c}_\ell^m(\rr,t),
    \label{eq:plm}
\end{equation}
with
\begin{equation}
    \label{eq:defclm}
    \mathbf{c}_\ell^m(\rr,t) \equiv \int \dd\nn \, \nn P(\rr,\nn;t) Y_\ell^{m}(\nn)^*.
\end{equation}
The expansion on spherical harmonics of $\nn$ is, 
\begin{equation}
    \nn = \sum_{\ell,m} \boldsymbol{\alpha}_\ell^m Y_\ell^m(\nn),
\end{equation}
with
\begin{equation}
    \boldsymbol{\alpha}_\ell^m = \sqrt{\frac{2\pi}{3}}\delta_{\ell,1} \left[ \delta_{m,-1}(\hat{\ee}_x + \ii \hat{\ee}_y) - \delta_{m,1}(\hat{\ee}_x - \ii \hat{\ee}_y) + \sqrt{2}\delta_{m,0}\hat{\ee}_z \right].
\end{equation}
Then, we replace $\nn$ and $P(\rr,\nn;t)$ by their expansions in Eq. \eqref{eq:defclm} and we use the orthogonality and the following contraction rule of spherical harmonics~\cite{Rose1957}: 
\begin{equation}
    Y_{\ell'}^{m'}(\nn)Y_{\ell''}^{m''}(\nn) = \sum_{L,M} \sqrt{\frac{(2\ell'+1)(2\ell''+1)}{4\pi(2L+1)}} \langle \ell' m' \ell'' m'' |LM\rangle Y_L^M(\nn).
\end{equation}
We find that
\begin{align}
    \mathbf{c}_\ell^m(\rr,t) &= \sum_{\ell',m'} \sqrt{\frac{(2\ell'+1)}{2(2\ell+1)}} p_{\ell'}^{m'}(\rr,t) \langle \ell' 0 10|\ell 0 \rangle \nonumber\\
    &\times \left[ \langle \ell' m' 1 -1 | \ell m \rangle(\hat{\ee}_x + \ii \hat{\ee}_y) - \langle \ell' m' 1 1 | \ell m \rangle (\hat{\ee}_x - \ii \hat{\ee}_y) + \sqrt{2}\langle \ell' m' 1 0 | \ell m \rangle \hat{\ee}_z  \right] .
    \label{eq:clm}
\end{align}
The coefficients $\langle \ell' m' \ell'' m'' |LM\rangle$ that appear in the expression are the Clebsch-Gordan coefficients, which are numbers that typically arise from angular momentum coupling in quantum mechanics. These coefficients have a rather complicated general expression, but most of them are vanishing. They can only be non-zero when, 
\begin{align}
    |\ell-\ell'| &< L < \ell + \ell' \\
    M &= m+m'
\end{align}
Note that these coefficients are related to the Wigner 3-j symbols by the relation, 
\begin{equation}
    \langle \ell m \ell' m|LM \rangle = (-1)^{-\ell+\ell'-M} \sqrt{2L+1} \begin{pmatrix}
        \ell & \ell' & L \\
        m & m' & -M
    \end{pmatrix}
\end{equation}
Eq. \eqref{eq:plm}, together with the expression of the coefficients $\mathbf{c}_\ell^m $ [Eq. \eqref{eq:clm}], gives the coupled set of equations obeyed by the expansion of the propagator $P$ on spherical harmonics.

We now relate explicitly this expansion to the moment expansion discussed in Section \ref{sec:FPE_momentexp_ani}. Up to order $\ell=2$ included, the expansion then reads:
\begin{eqnarray}
    P(\rr,\nn;t) &=&  Y_0^0(\nn) \int \dd \nn' \; P(\rr,\nn';t)Y_0^0(\nn')^* + \sum_{m=-1}^1 Y_1^m(\nn) \int \dd \nn' \; P(\rr,\nn';t) Y_1^m(\nn')^*  \nonumber\\
    &&+ \sum_{m=-2}^2 Y_1^m(\nn) \int \dd \nn' \; P(\rr,\nn';t) Y_1^m(\nn')^* + \dots
\end{eqnarray}
To make explicit calculations, the unit vectors $\nn$ are written in spherical coordinates: $\nn = (\sin\theta\cos\varphi,\sin\theta\sin\varphi,\cos\theta)$; $\nn' = (\sin\theta'\cos\varphi',\sin\theta'\sin\varphi',\cos\theta')$. We then use the expressions of the spherical harmonics of orders $\ell=0$, $1$ and $2$, that are recalled for completeness:
\begin{itemize}
    \item $\ell = 0$:
    \begin{equation}
         Y_0^0(\theta,\varphi) = \frac{1}{\sqrt{4\pi}}
    \end{equation}
    \item $\ell = 1$:
    \begin{equation}
        Y_1^{-1}(\theta,\varphi) = \sqrt{\frac{3}{8\pi}} \sin \theta\,  \ex{-\ii\varphi}
        ~;~ 
        Y_1^{0}(\theta,\varphi) = \sqrt{\frac{3}{4\pi}} \cos \theta 
        ~;~ 
        Y_1^{1}(\theta,\varphi) =- \sqrt{\frac{3}{8\pi}} \sin \theta\,  \ex{\ii\varphi}
    \end{equation}
    \item $\ell = 2$:
    \begin{align}
    &Y_2^{-2}(\theta,\varphi) = \sqrt{\frac{15}{16\pi}} \sin^2 \theta\,  \ex{-2\ii\varphi}
    ~;~
    Y_2^{-1}(\theta,\varphi) = \sqrt{\frac{15}{8\pi}} \sin \theta\cos\theta\,  \ex{-\ii\varphi}
    ~;~
    Y_2^{0}(\theta,\varphi) = \sqrt{\frac{5}{16\pi}} (3\cos^2 \theta-1)\nonumber\\
    &Y_2^{1}(\theta,\varphi) = -\sqrt{\frac{15}{8\pi}} \sin \theta\cos\theta\,  \ex{\ii\varphi}
    ~;~
    Y_2^{2}(\theta,\varphi) = \sqrt{\frac{15}{16\pi}} \sin^2 \theta\,  \ex{2\ii\varphi}
    \end{align}
\end{itemize}
We finally get:
\begin{eqnarray}
    P(\rr,\nn;t) &=&   \frac{1}{4\pi} \int \dd \nn' \; P(\rr,\nn',t)  +\frac{3}{4\pi} n_i \int \dd \nn' \; n'_i P(\rr,\nn';t) \nonumber\\
    && +\frac{15}{8\pi} \left(n_i n_j -\frac{1}{3} \delta_{ij} \right)  \int \dd \nn' \left(n'_i n'_j -\frac{1}{3}\delta_{ij} \right)P(\rr,\nn';t) +\dots \\
    &=& \frac{1}{4\pi} \rho(\rr,t) + \frac{3}{4\pi} n_i T_i(\rr,t) +\frac{15}{8\pi} \left(n_i n_j -\frac{1}{3} \delta_{ij} \right)  Q_{ij}(\rr,t) + \dots
\end{eqnarray}
This establishes the relationship between the moments $\rho$, $\boldsymbol{T}$ and $\mathbf{Q}$, and the expansion of $P$ over the spherical harmonics.

In order to retrieve the set of equations \eqref{eq:dens_ani} \eqref{eq:pola_ani} \eqref{eq:nema_ani}, one must use the expansion in spherical harmonics of $\nn$ and $\nn\nn$ to write $\TT$ and $\QQ$ as a function of the coefficients $p_\ell^m$. Due to function parity, an even rank tensor of the moment expansion can be expressed by the coefficients $p_\ell^m$ with an even $\ell$. Similarly, the odd rank tensor can be expressed with $p_\ell^m$ with odd $\ell$. In the following, the components of $\TT$ and $\QQ$ are given as functions of $p_\ell^m$. We finally get:
\begin{itemize}
    \item order $0$:
    \begin{equation}
        \rho = \sqrt{4\pi} p_0^0
    \end{equation}
    \item order $1$:
    \begin{align}
    &T_1 = \sqrt{\frac{2\pi}{3}}(p_1^{-1} - p_1^1)  \\
    &T_2 = -\ii \sqrt{\frac{2\pi}{3}}(p_1^{-1} + p_1^1)  \\
    &T_3 = \sqrt{\frac{4\pi}{3}}p_1^0 
    \end{align}
    \item order $2$:
    \begin{align}
    &Q_{11} = \sqrt{\frac{2\pi}{15}} \left( p_2^2 + p_2^{-2} - \sqrt{\frac{2}{3}} p_2^0 - \sqrt{\frac{10}{3}}p_0^0 \right)   \\
    &Q_{33} = \sqrt{\frac{2\pi}{15}} \left( \sqrt{\frac{8}{3}}p_2^0 - \sqrt{\frac{10}{3}}p_0^0 \right)  \\
    &Q_{12} = Q_{21} = \ii \sqrt{\frac{2\pi}{15}} \left( p_2^2 - p_2^{-2} \right)  \\
    &Q_{13} = Q_{31} = \sqrt{\frac{2\pi}{15}} \left(p_2^{-1} - p_2^1 \right)  \\
    &Q_{23} = Q_{32} = \sqrt{\frac{2\pi}{15}} \left( p_2^1 + p_2^{-1} \right)  \\
    &Q_{22} = -Q_{11} - Q_{33}
    \end{align}
\end{itemize}

\section{Resolution of the equations closed at order 2}
\label{app:solution_closure_order_2}

\begin{figure}
\begin{center}
\includegraphics[width= 0.8\textwidth]{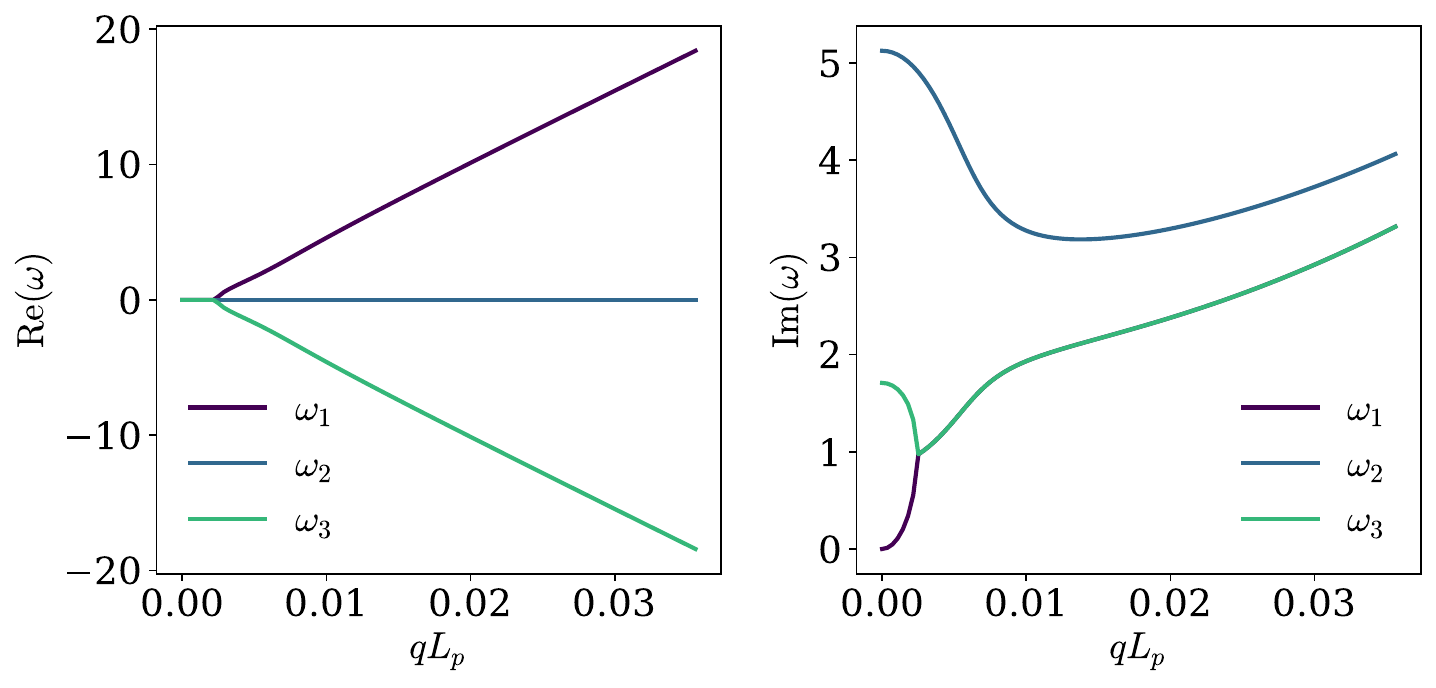}
\caption{{Left:} Real and {Right:} imaginary part of the poles of $\text{det}(\mathbf{A})$, defined in Eq. \ref{eq:def_A}. Parameters: $P_e$ = 20.1, $\Delta D/\bar{D}_t =0.2$,  $a/b = 3$, $L_p = 28.1$.}
\label{fig:poles_order2}
\end{center}
\end{figure}

In this Appendix, we solve Eqs.~\eqref{eq:closure2_1}-\eqref{eq:closure2_3}, which is a linear set of equations satisfied by $\widetilde{\rho}$,${T}_{\parallel}$ and ${Q}_{\parallel}$. It can be rewritten in matrix form:
\begin{equation}
    \boldsymbol{A} \boldsymbol{X} = \boldsymbol{S}
\end{equation}
with, 
\begin{equation}
A = \begin{bmatrix}
(\ii\omega + D_0q^2) & \ii vq & \Delta Dq^2 \\
\frac{1}{d}\ii vq & \left(\ii\omega + (D_1 + \frac{2}{d+2}\Delta D)q^2 + (d-1)D_r\right) & \ii vq \\
\frac{2(d-1)}{d^2 (d+2)}\Delta Dq^2 & \frac{2(d-1)}{d(d+2)} \ii vq & \left(\ii\omega + (D_2 + \frac{4(d-1)}{d(d+4)}\Delta D)q^2 + 2dD_r\right)
\end{bmatrix}
\label{eq:def_A}
\end{equation}
and, 
\begin{equation}
\boldsymbol{X} = \begin{bmatrix}
\tf{\rho}_2 \\
{T}_{\parallel} \\
{Q}_{\parallel}
\end{bmatrix}
\quad ; \quad \boldsymbol{S} = \begin{bmatrix}
1 \\
0 \\
0
\end{bmatrix}.
\end{equation}
Thus, we have, 
\begin{equation}
    \tf{\rho}(q;\omega) = (\boldsymbol{A}^{-1})_{1,1}.
\end{equation}
Using the adjugate matrix we can write, 
\begin{equation}
    \boldsymbol{A}^{-1} = \text{det}(\boldsymbol{A})^{-1} \text{adj}(\boldsymbol{A}),
\end{equation}
with $\text{adj}(\boldsymbol{A})_{i,j} = (-1)^{i+j}\text{det}(\boldsymbol{M}_{j,i})$, and where $\boldsymbol{M}_{j,i}$ is the matrix that results from removing the row $j$ and the column $i$ from $\boldsymbol{A}$. Thus, one gets
\begin{equation}
    \tf{\rho}(q;\omega) = \text{det}(\boldsymbol{A})^{-1} \text{det}(\boldsymbol{M}_{1,1})
\end{equation}
In order to get the ISF we need to perform a temporal inverse Fourier transform. As in Section \ref{sec:closure1}, we will use the residue theorem, and thus, and we need the poles of $\text{det}(\boldsymbol{A})$ (which is a degree 3 polynomial in $\omega$). We can write this polynomial as 
\begin{equation}
    \text{det}(\mathbf{A}) = a\omega^3 + b\omega^2 + c\omega + d
\end{equation}
with (for $d=3$), 
\begin{align}
    a &= -\ii, \\
    b &= -\frac{1}{35}(54D_{\perp} + 51D_{\parallel})q^2 - 8D_r, \\
    c &= \ii \left( \frac{12}{7} D_{\perp}D_{\parallel} + \frac{3}{5} D_{\parallel}^2 + \frac{24}{35}D_{\perp}^2 \right) q^4 + \ii \left( \frac{256}{35}D_{\parallel}D_r + \frac{304}{35}D_{\perp}D_r + \frac{3}{5}v^2 \right) q^2 + 12 \ii D_r^2, \\
    d &= \frac{1}{175}( 72 D_{\parallel}D_{\perp}^2 + 78 D_{\parallel}^2 D_{\perp} + 9 D_{\parallel}^3 + 16 D_{\perp}^3 )q^6 , \\
    &+ \left( \frac{72}{35}D_{\perp}^2 D_r + \frac{18}{35}D_{\perp}v^2 + \frac{48}{35} D_{\parallel}^2 D_r + \frac{3}{35}D_{\parallel}v^2 + \frac{32}{7}D_{\parallel}D_{\perp}D_r \right)q^4
    + (4 D_{\parallel}D_r^2 + 8 D_{\perp}D_r^2 + 2 D_e v^2)q^2 .
\end{align}
Thus, the poles read 
\begin{equation}
    \omega_k = -\frac{1}{3a} \left( b+j^k C + \frac{\Delta_0}{j^k C} \right)
\end{equation}
with
\begin{align}
    &C = \left( \Delta_1 \pm \sqrt{\Delta_1^2 - 4\Delta_0^3} \right)^{1/3} \nonumber \\
    &\Delta_0 = b^2 - 3ac \nonumber \\
    &\Delta_1 = 2b^3 - 9abc + 27a^2d \nonumber \\
    & j = \exp{\left(\ii \frac{2\pi}{3}k\right)}, \quad k \in \{0,1,2\}
\end{align}
We take as a contour for the residue theorem a half disk in the $\mathbb{R}_+$ plane. Because the imaginary part of the three poles is always positive, they always are in the contour (see Fig.\ref{fig:poles_order2}), and therefore the ISF at order 2 and the first and second moments projected on $q$ are,
\begin{equation}
    F_2(q,t) = \int \frac{\dd \omega}{2\pi} \widetilde{\rho}(q,\omega)\ex{\ii \omega t}= \sum_{i=1}^3 \prod_{j \neq i} \frac{P_1(\omega_i)\ex{\ii \omega_i t}}{(\omega_i - \omega_j)}
\end{equation}
\begin{equation}
    T_{\parallel,2}(q,t) = \sum_{i=1}^3 \prod_{j \neq i} \frac{\mathcal{P}_2(\omega_i)\ex{\ii \omega_i t}}{(\omega_i - \omega_j)}
\end{equation}
\begin{equation}
    Q_{\parallel,2}(q,t)  = \sum_{i=1}^3 \prod_{j \neq i} \frac{\mathcal{P}_3(\omega_i)\ex{\ii \omega_i t}}{(\omega_i - \omega_j)}
\end{equation}
with,
\begin{align}
    \mathcal{P}_i(\omega) &=(-1)^{i}\text{det}(\mathbf{M}_{1,i}) \nonumber \\
    \mathcal{P}_1(\omega)&= \omega^2 - \ii \left( \frac{92}{105} D_{\perp}q^2 \frac{118}{105}D_{\parallel}q^2 + 8 D_r \right)\omega \nonumber \\
    &- \left( \frac{52}{105}D_{\parallel}D_{\perp}q^4 + \frac{11}{35}D_{\parallel}^2 q^4 + \frac{4}{21}D_{\perp}^2 q^4 + \frac{488}{105}D_{\parallel}D_rq^2 \frac{352}{105}D_{\perp}D_r q^2 + 12D_r^2 + \frac{4}{15}v^2 q^2 \right) \\
    \mathcal{P}_2(\omega) &= -\frac{vq}{3}\omega + \frac{\ii v q^3}{105} \left(26D_\perp + 9D_\parallel \right) + 2\ii vq D_r  \\
    \mathcal{P}_3(\omega) &= \frac{4\ii}{45}  (D_{\parallel}-D_\perp)q^2\omega  - \frac{4}{225} D_\parallel D_\perp q^4 - \frac{8}{225}D_\perp^2 q^4 + \frac{4}{75}D_\parallel^2 q^4 + \frac{8}{45}D_r (D_\parallel-D_\perp)q^2 + \frac{4}{45}v^2q^2.
\end{align}
These are Eqs.~\eqref{eq:F2}-\eqref{eq:P3} from the main text.
 
\section{Note on multiplicative noise}
\label{app:multiplicativenoise}

In this Appendix, we discuss the discretization scheme of the Langevin equation obeyed by orientation [Eq. \eqref{eq:Langevin_ori0} in the main text], that we recall here for completeness:
\begin{equation}
    \frac{\dd \nn}{\dd t}(t) =  \sqrt{2 D_r}\boldsymbol{\zeta}(t)  \times \nn(t)
    \label{eq:Langevin_ori_app}
\end{equation}
There is no particular remark to be made in 2D, as this equation can be integrated straightforwardly with a usual Euler scheme. However, in 3D, the difficulty in the discretization lies in the nature of the noise term, whose amplitude depends on $\nn$, and which is therefore multiplicative. In addition, the norm of $\nn$ must be conserved and equal to 1 at all times: this is encoded in Eq. (\ref{eq:Langevin_ori_app}) by the cross product term and can be easily checked by taking the scalar product by $\nn$:
\begin{equation}
    \nn \cdot \frac{\dd \nn}{\dd t} = \frac{1}{2} \frac{\dd}{\dd t}[\nn^2] = 0,
\end{equation}
meaning that, $\vert \nn(t) \vert = \vert \nn(0) \vert = 1.$ Therefore, the discretization scheme must conserve this property.

We start by integrating the $i$-th component of Eq. (\ref{eq:Langevin_ori_app}) on a small time interval $\Delta t$,
\begin{equation}
    \Delta n_i(t) = n_i(t+\Delta t)-n_i(t)= - \sqrt{2D_r} \epsilon_{ijk} \int_{t}^{t+\Delta t} \dd t' n_j(t') \zeta_k(t'),
    \label{eq:Int_Langevin_ori}
\end{equation}
where $\boldsymbol{\epsilon}$ is the Levi-Civita tensor. We then write
\begin{align}
     n_j(t') &=  n_j(t) + \Delta n_j(t') + ...
\end{align}
Inserting this in Eq. (\ref{eq:Int_Langevin_ori}) and replacing $\dd n_j(t')$ by its expression, we get
\begin{equation}
    \Delta n_i(t) = -\sqrt{2D_r} \epsilon_{ijk} n_j(t) \Delta W_k(t) -\sqrt{2D_r} \epsilon_{ijk} \int_{t}^{t+\Delta t} \dd t' \Delta n_j(t') \zeta_k(t') + \text{O}(\Delta t^{\frac{3}{2}})
\end{equation}
where $\boldsymbol{W}(t)$ is a Wiener process, i.e. such that $\Delta W_i(t) = \int_{t}^{t+\Delta t}\dd t' \zeta_i(t')$. We also recall here that, 
\begin{align}
    &\langle \Delta W_i(t) \rangle = 0 \\
    &\langle \Delta W_i(t) \Delta W_j(t) \rangle = \delta_{ij} \Delta t \nonumber    
\end{align}.
Then we insert Eq. (\ref{eq:Int_Langevin_ori}) again in this expression to get, 
\begin{align}
    \Delta n_i(t) &= -\sqrt{2D_r} \epsilon_{ijk} n_j(t) \Delta W_k(t) -\sqrt{2D_r} \epsilon_{ijk} \int_{t}^{t+\Delta t} \dd t' \left( -\sqrt{2D_r} \epsilon_{jpq} \int_{t}^{t'} \dd t'' n_p(t'')\zeta_q(t'') \right) \zeta_k(t') + \text{O}(\Delta t^{\frac{3}{2}})\\
    &= -\sqrt{2D_r} \epsilon_{ijk} n_j(t) \Delta W_k(t) + 2D_r \epsilon_{ijk}\epsilon_{jpq} \int_{t}^{t+\Delta t} \dd t' \int_{t}^{t'} \dd t'' n_p(t'')\zeta_q(t'') \zeta_k(t') + \text{O}( \Delta t^{\frac{3}{2}})
\end{align}
We Taylor expand $n_p(t'')$ at order 0 (because the order 1 term would give a $O(dt^{\frac{3}{2}})$ term): 
\begin{equation}
    \Delta n_i(t) = -\sqrt{2D_r} \epsilon_{ijk} n_j(t) \Delta W_k(t) + 2D_r \epsilon_{ijk}\epsilon_{jpq} n_p(t) \int_{t}^{t+\Delta t} \dd t' \int_{t}^{t'} \dd t'' \zeta_q(t'') \zeta_k(t') + \text{O}(\Delta t^{\frac{3}{2}})
\end{equation}
and using the following property of Levi-Civita tensor: $\epsilon_{ijk}\epsilon_{jpq} = \delta_{kp} \delta_{iq} - \delta_{kq} \delta_{ip}$, we get 
\begin{eqnarray}
    \Delta n_i(t) &=& -\sqrt{2D_r} \epsilon_{ijk} n_j(t) \Delta W_k(t) \nonumber\\
    &&+ 2D_r \left[ n_k(t) \int_{t}^{t+\Delta t} \dd t' \int_{t}^{t'} \dd t'' \zeta_i(t'') \zeta_k(t') - n_i(t) \int_{t}^{t+\Delta t} \dd t' \int_{t}^{t'} \dd t'' \zeta_k(t'') \zeta_k(t') \right] + \text{O}(\Delta t^{\frac{3}{2}})   
    \label{eq:dn}
\end{eqnarray}
We now search for the expression of the first and second cumulant of $\Delta \nn(t)$. 
\begin{align}
    \moy{\Delta n_i(t)} &= 2D_r \left[ n_k(t) \int_{t}^{t+\Delta t} \dd t' \int_{t}^{t'} \dd t'' \delta_{ik} \delta(t'-t'') - n_i(t) \int_{t}^{t+\Delta t} \dd t' \int_{t}^{t'} \dd t'' \delta_{kk} \delta(t'-t'') \right] + \text{O}(\Delta t^{\frac{3}{2}})\\
    &= - 4D_r n_i(t) \int_{t}^{t+\Delta t} \dd t' \int_{t}^{t'} \dd t'' \delta(t'-t'')  + \text{O}(\Delta t^{\frac{3}{2}})
\end{align}

In this formulation, the choice between the Itô and Stratonovitch conventions is converted into a choice on the value of the integral $\int_{t}^{t'} \dd t'' \delta(t'-t'')$, which is ambiguous since $t'$ is present both in the argument of the $\delta$-function and in the bound of the integral. Relying on Ref.~\cite{Gardiner1985}, for an arbitrary function $f$, we define the integrals:
\begin{equation}
    \begin{cases}
    I_1 = \int_{\tau_1}^{\tau_2} \dd \tau \, f(\tau)\delta(\tau-\tau_1) \\
I_2 = \int_{\tau_1}^{\tau_2} \dd \tau \, f(\tau)\delta(\tau-\tau_2) \\
    \end{cases}
\end{equation}
which read, in the respective conventions:
\begin{equation}
\text{(It\^o)}~
    \begin{cases}
    I_1 = f(t_1) \\
I_2 =  0\\
    \end{cases}
    \qquad;\qquad
       \text{(Stratonovitch)}~ \begin{cases}
    I_1 = \frac{1}{2}f(t_1) \\
I_2 =  \frac{1}{2}f(t_2)\\
    \end{cases}
\end{equation}
Here, taking $f(t)=1$, we thus have, 
\begin{equation}
    \langle \Delta n_i(t) \rangle = -4 D_r I_2 n_i(t) \Delta t + \text{O}(\Delta t^{\frac{3}{2}}).
\end{equation}
The covariance matrix of $\Delta \nn(t)$ is, 
\begin{align}
    \moy{\Delta n_i \Delta n_j} &= 2D_r  \epsilon_{iab} \epsilon_{jcd} n_a(t) n_c(t) \langle \Delta W_b(t) \Delta W_d(t) \rangle + \text{O}(\Delta t^{\frac{3}{2}})  \\
    &= 2 D_r \epsilon_{iab} \epsilon_{jcb} n_a(t) n_c(t) \Delta t + \text{O}(dt^{\frac{3}{2}}) + \text{O}(\Delta t^{\frac{3}{2}}) \nonumber \\
    &= 2D_r (\delta_{ij} - n_i n_j) \Delta t + \text{O}(\Delta t^{\frac{3}{2}}) \nonumber.
\end{align}
One can verify that all higher cumulants are of order $\text{O}(\Delta t^{3/2})$ or higher. Therefore, we can sample the random variable $\Delta \nn(t)$ with a normal distribution of average $\langle \Delta n_i \rangle$ and covariance $\langle \Delta n_i \Delta n_j \rangle$. And we get the following discretization schemes:
\begin{eqnarray}
    \text{(It\^o)}\qquad &&n_i(t+\Delta t) = n_i(t)  - \sqrt{2D_r} \epsilon_{ijk} n_j(t) \mathcal{N}_k(0,1) \sqrt{\Delta t}. \\
     \text{(Stratonovitch)}\qquad &&n_i(t+\Delta t) = n_i(t) - 2 D_r n_i(t) \Delta t - \sqrt{2D_r} \epsilon_{ijk} n_j(t) \mathcal{N}_k(0,1) \sqrt{\Delta t} \label{eq:discretization_Strato}
\end{eqnarray}
with $\mathcal{N}_k(0,1)$ a standard normal distribution. And one can verify that the norm is conserved only with the Stratonovich convention. We adopt Eq. \eqref{eq:discretization_Strato} to simulate anisotropic ABP in 3D.

\section{Influence of the aspect ratio of the particle}
\label{app:aspect_ratio}

\begin{figure}
\begin{center}
\includegraphics[width= 0.8\textwidth]{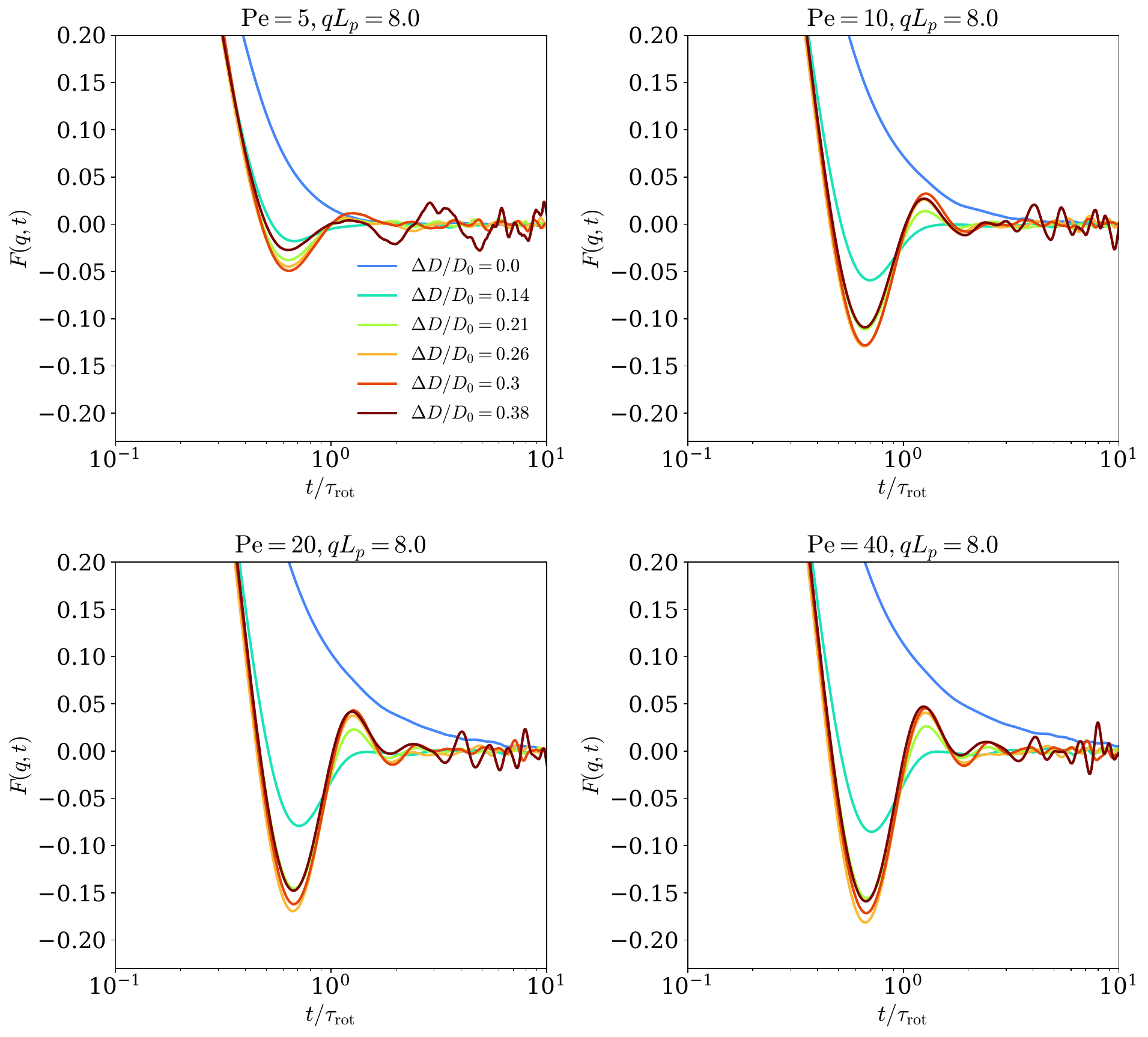}
\caption{ISF measured in numerical simulations as a function of time for different anisotropic parameters and for a given wavenumber. The P\'eclet number is kept constant by changing the self-propulsion velocity. Parameters: aspect ratio $a/b = [1,2, 3, 4, 5, 10]$, $\tau_{\text{rot}} = [0.17,0.50,1.17,2.26,3.87,22.3]$.}
\label{fig:isf_deltaD}
\end{center}
\end{figure}

Here we discuss the influence of the anisotropy on the ISF. We show in Fig. \ref{fig:isf_deltaD} the ISF for different anisotropic parameters, P\'eclet numbers and a given wavevector (chosen to be at an intermediate value where we can observe the signature of activity). As the P\'eclet number is a function of the anisotropy, we keep it constant by changing the self-propulsion velocity accordingly. We observe that the signature of the activity stays at the same time, and that the period of the oscillations is unchanged with a change in the anisotropy. As the signature is larger for a larger aspect ratio, it means that increasing the anisotropy increases the deviation from the Gaussian behavior. Formally, it comes from the fact that the anisotropy couples the positional probability distribution through the nematic tensor [Eq.~\eqref{eq:dens_ani}].

\begin{figure}
\begin{center}
\includegraphics[width= 0.8\textwidth]{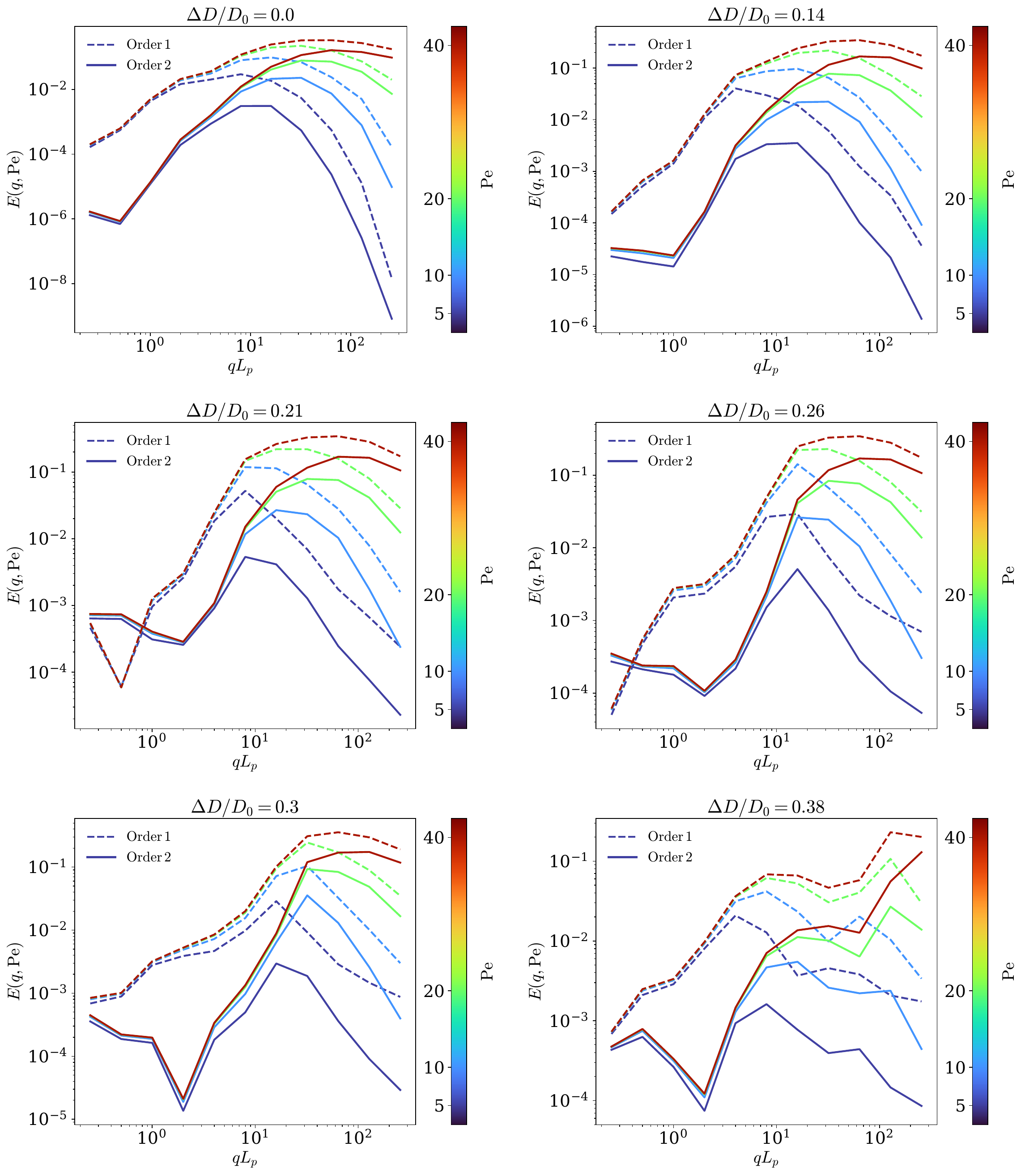}
\caption{Error between the ISF estimated from closures at orders 1 and 2 and numerical simulations, as defined in Eq. \eqref{eq:error}, and as a function of the rescaled wavevector $q L_p$, for different values of $\Pe$ and for different values of the anisotropic parameter. Parameters: aspect ratio  $a/b = [1,2,3,4,5,10]$.}
\label{fig:error_deltaD}
\end{center}
\end{figure}

We show in Fig. \ref{fig:error_deltaD} the error between the exact solution of the ISF and the ISF given by the closures at order 1 and 2 for different anisotropic parameters. One can notice that the validity of the different approximations is not influenced much by a change in anisotropy at a constant P\'eclet number, since the behavior of the error is qualitatively similar. Globally, the closure at order 2 is almost always better. At small values of $qL_p$ both orders perform well, but order 2 is more stable as $qL_p$ increases. It has a minimum at small values of $qL_p$ that is located at a larger $qL_p$ as the anisotropy increases. For small anisotropies, the error is smaller, but as the anisotropy is increased, its value stays at the same order of magnitude. For intermediate values of $qL_p$ the error is maximal as it is where the signature of the activity is larger. Finally, at large values of $qL_p$, the error decreases as an effective diffusive regime is retrieved at this scale.

\bibliography{biblio}

\end{document}